\documentclass[twocolumn]{mn2e}
\usepackage{graphics}
\title[Light curve solutions for eclipsing binaries in SMC]{Light curve 
solutions for bright detached eclipsing binaries in SMC: 
absolute dimensions and distance indicators}
\author[D.~Graczyk]
           {Dariusz Graczyk \thanks{e-mail:weganin@astri.uni.torun.pl}\\
            Institute of Astronomy, Lubuska 2, 65-265 Zielona G{\'o}ra, Poland\\
					Centre for Astronomy, Nicolaus Copernicus University, 
            Gagarina 11, 87-100 Toru{\'n}, Poland}           
\begin{document}
\date{}

\pagerange{01-14}
\pubyear{2003}

\maketitle

\label{firstpage}

\begin{abstract}
This paper presents a careful and detailed light curve analysis of 
bright detached eclipsing binaries (DEB) in the Small Magellanic Cloud, 
discovered by OGLE collaboration, on the basis of recently available  
difference image analysis (DIA) photometry. The 19 binaries brighter than 16.4 mag in $I$ band and with the depth of 
primary and secondary eclipse greater than 0.25 mag were investigated. The solutions 
were obtained by a modified version of the Wilson-Devinney program. 
%It is shown that the new DIA photometry of eclipsing binaries in SMC allows 
%for much better determination of photometric parameters than the DoPHOT 
%photometry from old OGLE catalog (Udalski et al. 1998). 
The quality of DIA 
light curves -- a good phase coverage and relatively small scatter -- is 
enough to calculate realistic estimates for the third light $l_3$ and 
the argument of periastron $\omega_o$. It was found that solutions of detached, 
eccentric systems with flat light curve between eclipses usually may suffer from
indetermination of $l_3$ in contrast to those of similar systems having some proximity 
effects. 

The physical properties of the stars were estimated on the basis of their
photometric elements and indices assuming the distance modulus
to SMC $\sim 18.9$ and consistency between an empirical mass-luminosity 
relation and the flux scaling. The method was
tested on three LMC stars of known absolute dimensions and a good
agreement was found for $m-M \sim 18.5$. Such an approach may give fast and accurate 
estimates of absolute dimensions for large and homogeneous samples
of eclipsing binaries in the Magellanic Clouds and other close
galaxies. Moreover, this method allows also for independent estimation of \mbox{$E(B-V)$}
in the direction to a particular binary.  
The subset of six bright DEB's worth future intensive investigations
as likely distance indicators to SMC, was chosen.
They are SC3 139376, SC4 53898, SC5 129441, SC6
67221, SC6 215965 and SC9 175336.
\end{abstract}

\begin{keywords}
binaries: eclipsing -- Magellanic Clouds 
\end{keywords}

\section{Introduction}
%-----------------------------------
Accurate determination of absolute dimensions of eclipsing
binaries based, among others, on detailed light curve analysis is still a very important task 
of modern astronomy. It gives an 
opportunity for testing advanced evolutionary models of stars (e.g.~Pols
et al.~1997) 
but also allows for very precise distance determination. 
The method, at least in principle,
is quite straightforward (see~Paczy{\'n}ski 1997
for an outline and Clausen~2000 for a review). 
Three distance determination for eclipsing binaries in the Large Magellanic Cloud 
were quite recently presented: HV 2274 (Guinan et al.~1998), HV 982 (Fitzpatrick et 
al.~2001) and EROS 1044 (Ribas et al.~2002) giving the distance modulus $m-M
\sim 18.4$. 
Light curve solutions together with estimation of physical parameters 
and the distance to the Small Magellanic Cloud were obtained also for two eclipsing
binaries: HV 2226 (Bell et al.~1991) and HV 1620 (Pritchard et al.~1998) 
giving for the distance modulus $m-M \sim 18.6$. Very recently  
ten eclipsing binaries have been analysed in the central part of the SMC 
giving mean distance modulus $m-M=18.9\pm0.1$ (Harries, Hilditch \& Howarth 2003).
The light curve analysis in their paper was based on DoPHOT OGLE photometry
(Udalski et al.~1998).

In recent years, extensive CCD photometry have been undertaken during the
projects of microlensing searching toward the Magellanic Clouds and Galactic Bulge. 
As a byproduct, thousands of eclipsing binaries were discovered and placed in 
catalogs (Grison et al.~1995, Alcock et al.~1997, Udalski et al.~1998). 
Preliminary solutions for MACHO LMC catalog were found
by Alcock et al.~(1997). They fitted 611 light curves of eclipsing binaries using 
the EBOP code (Etzel 1993). That code uses a tri-axial ellipsoid 
approximation what makes this code very fast. But the code cannot properly 
account for the proximity effects present in most of the
eclipsing binaries detected by MACHO. For SMC eclipsing binaries OGLE 
collaboration (Udalski et al.~1998) extracted from their catalog a sample 
of 153 detached binaries with well defined, narrow eclipses of similar depth 
and relatively good photometry. It was argued that the sample contains 
the best systems for distance determination. 
Wyithe \& Wilson (2001, 2002 -- hereafter WW1, WW2) 
presented an excellent analysis of the whole OGLE data containing 1459 
eclipsing binaries using an automated version of the Wilson-Devinney (1971, 
Wilson 1992) code. This program can provide very detailed treatment of 
proximity effects. They used public 
domain DoPHOT photometry available at OGLE homepage 
{\it http://sirius.astrouw.edu.pl/ogle/ogle2/var$_{-}$stars/smc/ecl} in the analysis. 
They extracted in an objective way two subsets of 
detached binaries with very narrow eclipses and complete eclipses 
as the most likely good distance indicators. Subset of 
semi-detached eclipsing binaries was also chosen as distance indicators on the 
basis of similar surface brightness between components and complete 
eclipses. The authors intend, also, to extract such 
subsets for contact binaries.
In general it is believed that the close-to-ideal 
system chosen as a distance indicator, should have two stars of similar 
temperature and luminosity and the separation of the components should 
be significantly larger than their radii. But well-detached, eccentric systems 
may suffer from the aliasing of photometric solutions. 
WW2 argued that semi-detached 
and contact binaries, and also detached eclipsing binaries
showing larger proximity effects, have their photometric solutions 
better defined and may serve as potential best distance indicators as in the case
of recently analyzed 15-th magnitude binary EROS 1044 (Ribas et al.~2002). 
In this paper I focused my attention only on detached eclipsing binaries.

The release of new DIA photometry by OGLE collaboration motivated me to 
reanalyse the light curves of eclipsing binaries in SMC for the
purpose of extracting a sample of the best distance indicators. This sample 
was chosen using additional important criteria which were 
not used in the selections made by Wyithe \& Wilson. The analysis was restricted 
only for bright systems. It is obvious that those systems have, in general, 
the largest probability of obtaining high-quality light curves and radial 
velocity curves. Thus the searching  for distance indicators
should start from the analysis of the brightest candidates. Moreover using DIA 
photometry we can achieve more realistic determination of the light curve 
parameters than in the case of using the older DoPHOT photometry.

\section{The data}
%------------------------------------------
\subsection{Photometry}
%-------------------------
CCD differential photometric observations of SMC eclipsing binaries were 
obtained by OGLE between 1997 and 2000. The data were acquired using 1.3 m 
Warsaw Telescope at Las Campanas Observatory. The photometric $I$ indices were
calculated using the technique of difference image analysis (DIA) as was 
reported in Zebru{\'n}, Soszy{\'n}ski \& Wo{\'z}niak (2001a). The new 
photometry was quite recently facilitated on OGLE homepage 
{\it http://sirius.astrouw.edu.pl/ogle/ogle2/dia} (Zebru{\'n} 
et al.~2001b). The OGLE database contains all variables detected in the LMC and SMC
(68 thousand) during the course of the OGLE-II project. The DIA is argued as a 
superior method over classical DoPHOT method because of its better accuracy, 
especially in crowded fields (Zebru{\'n} et al.~2001a). 
The light curves based on OGLE-DIA photometry are characterized by considerably smaller 
scatter of observations and higher smoothness of the light curves. Also, in most cases
the phase coverage is better because the DIA photometry includes OGLE observations 
which have extended almost two years after the moment when original catalog of SMC 
eclipsing binaries (Udalski et al.~1998) was published. 

\begin{figure}
\begin{minipage}{\linewidth}
  \resizebox{8.8cm}{!}{\includegraphics{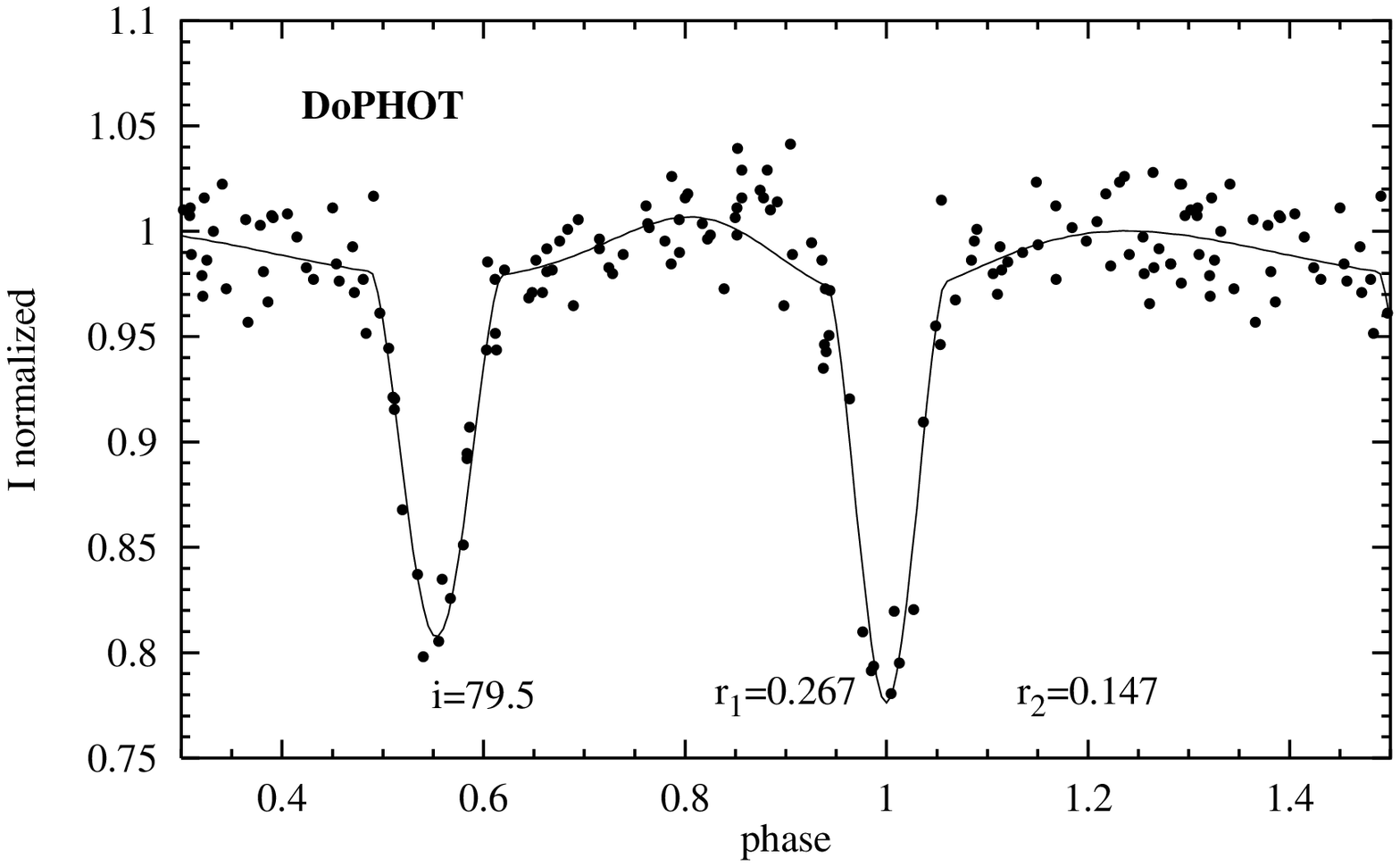}}
  \resizebox{8.8cm}{!}{\includegraphics{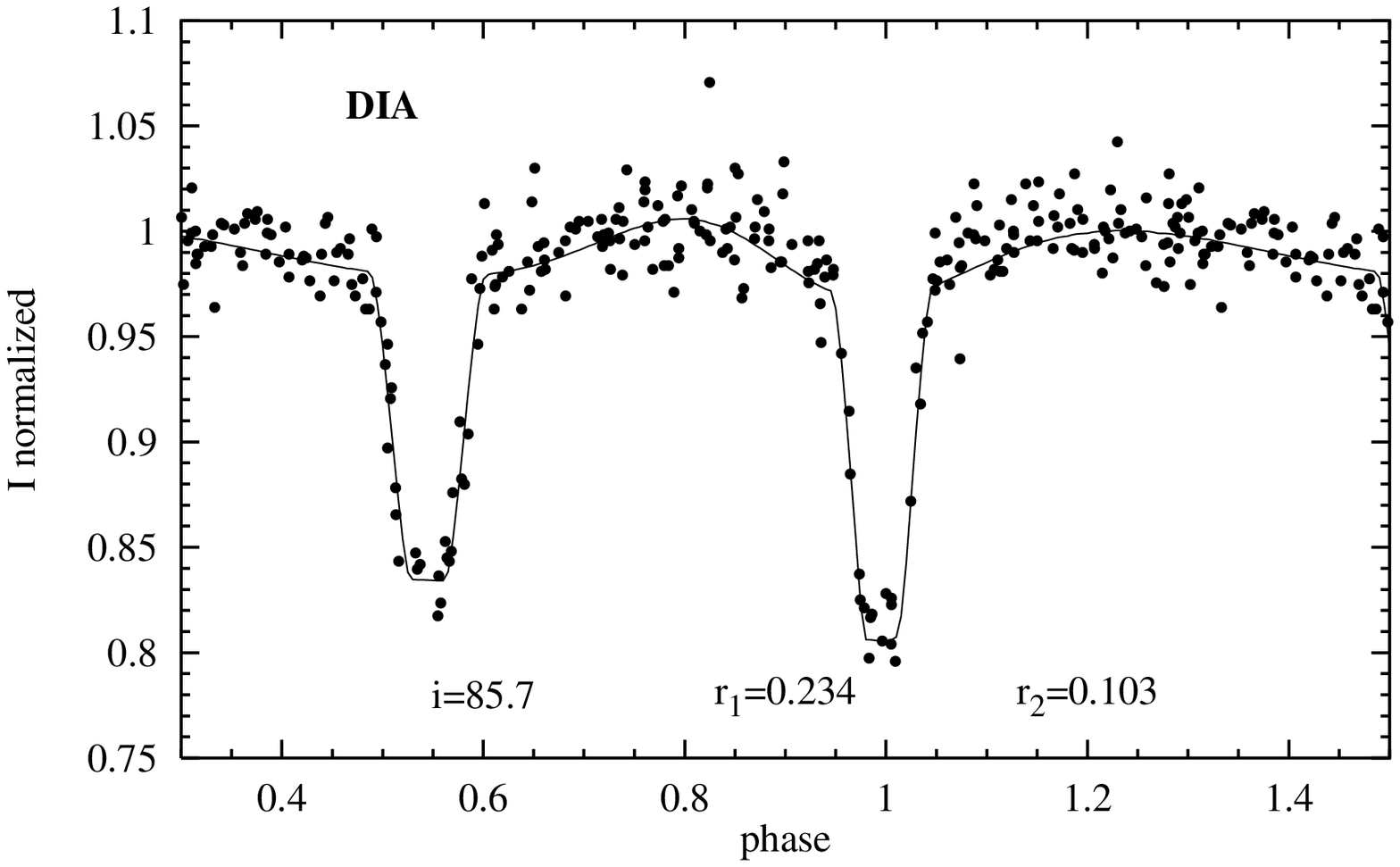}}
\end{minipage}
\caption{The light curves of SMC SC1 25589 eclipsing binary based on OGLE DoPHOT
photometry (upper panel) and OGLE DIA photometry (lower panel). Preliminary solutions
were marked by solid lines.}
\label{fig:25589}
\end{figure}

The quality of OGLE-DIA photometry can be demonstrated by a comparison of the 
$I$ light curves obtained with both methods. As an example I 
present a 17-th magnitude eccentric eclipsing binary SMC SC1 25589 -- 
Fig.~\ref{fig:25589} -- together with preliminary light curve solutions. 
Note that according to the analysis of DIA light curve the system has 
apparently total eclipses and much larger separation between components 
than in the case of the analysis of DoPHOT data.

\begin{figure}
\begin{minipage}{0.92\linewidth}
  \resizebox{\linewidth}{!}{\includegraphics{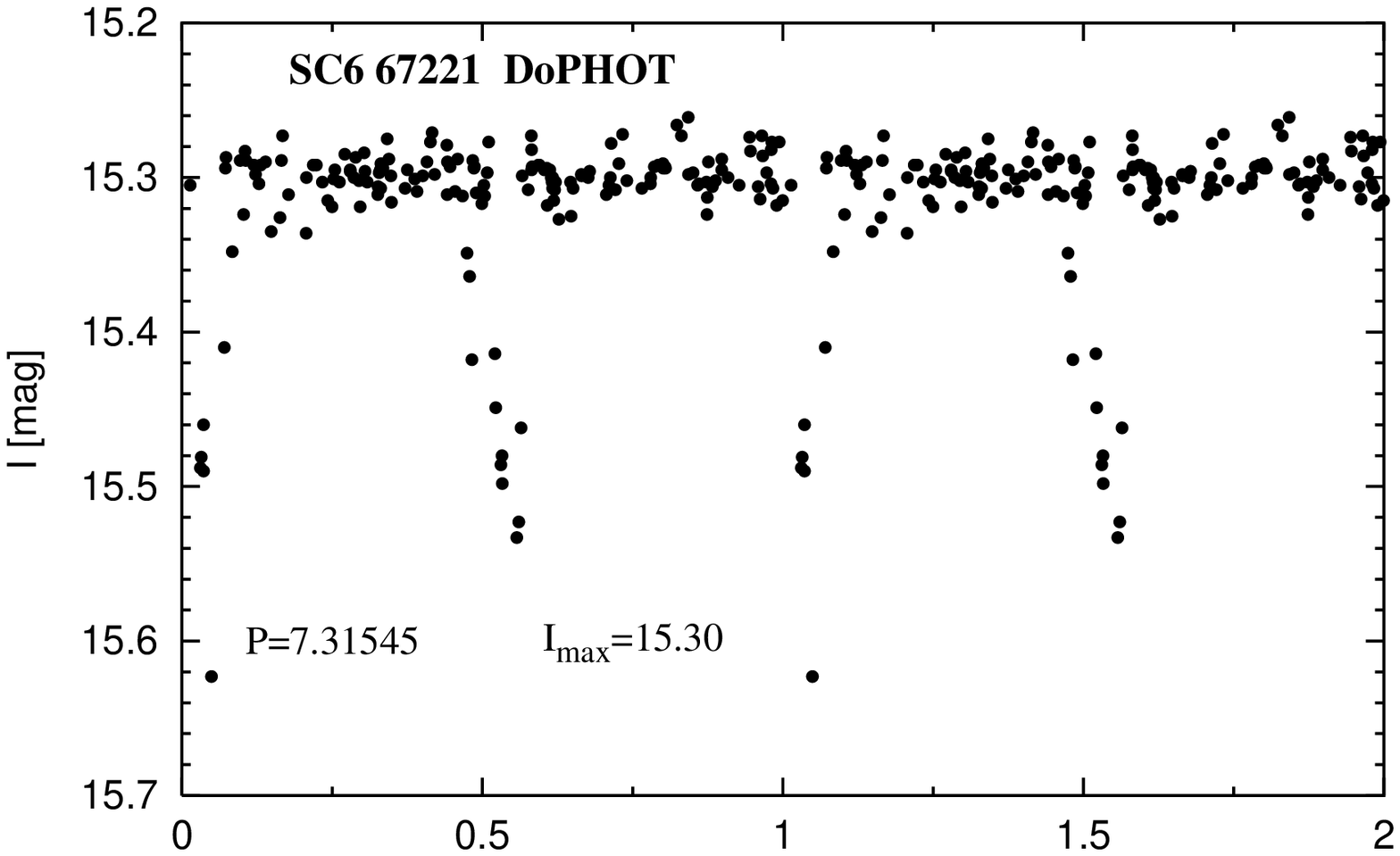}}
  \resizebox{\linewidth}{!}{\includegraphics{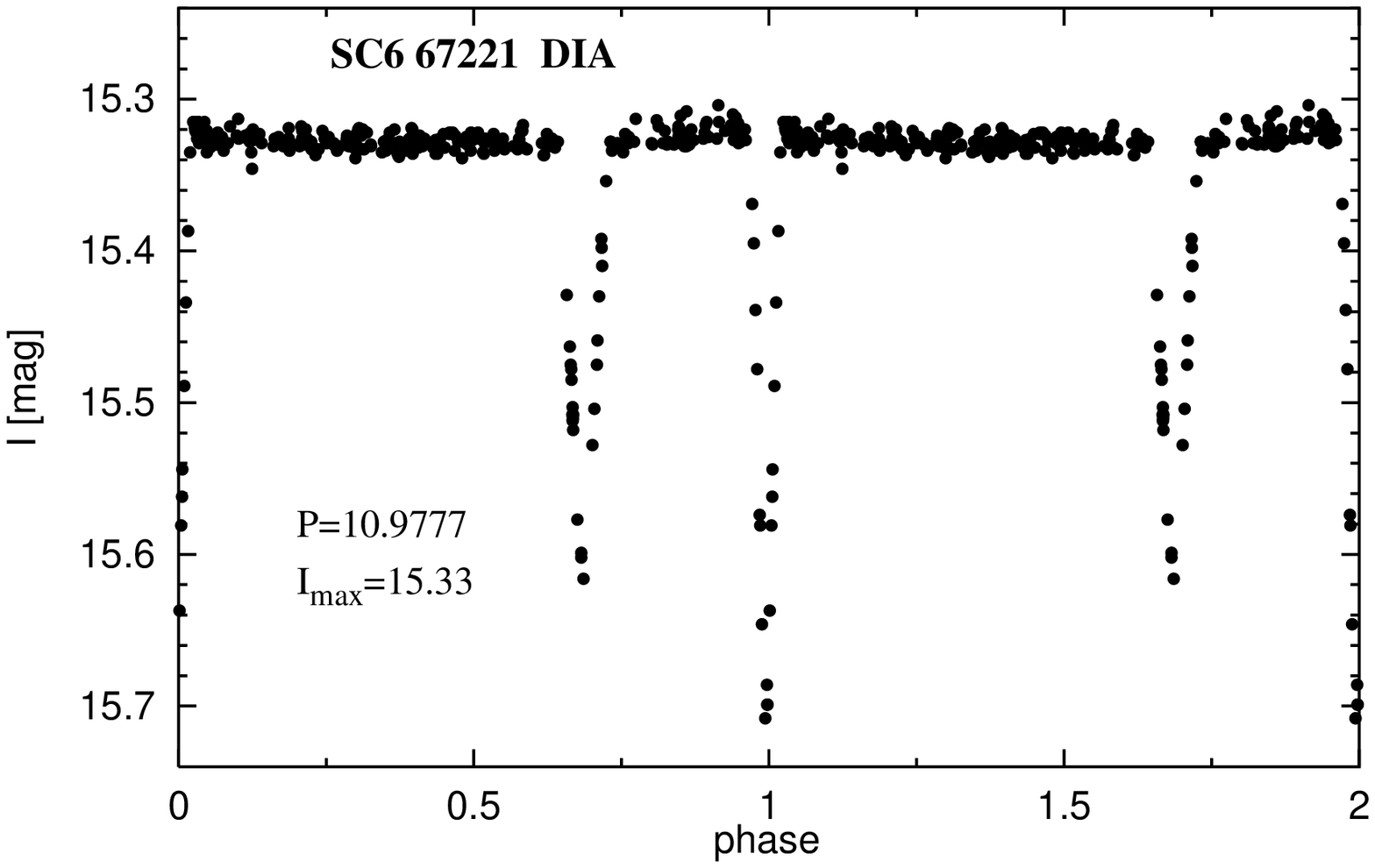}}
\end{minipage}
\caption{The light curves of SMC SC6 67221: from Udalski et al.~(1998) (upper panel)
and from the analysis of OGLE-DIA photometry in this paper (lower panel).}
\label{fig:67221com}
\end{figure}

\begin{figure}
\begin{minipage}{0.92\linewidth}
  \resizebox{\linewidth}{!}{\includegraphics{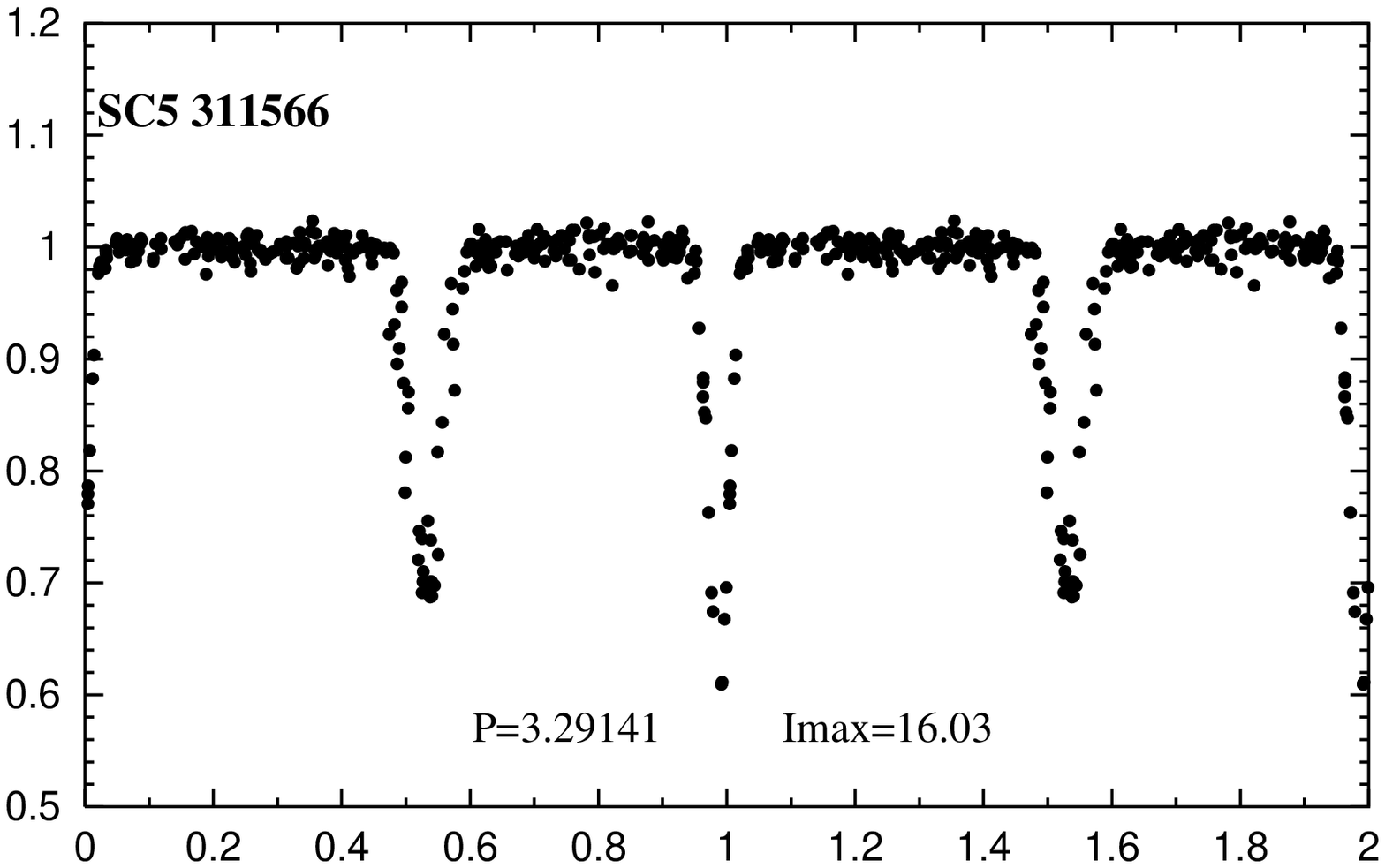}}
  \resizebox{\linewidth}{!}{\includegraphics{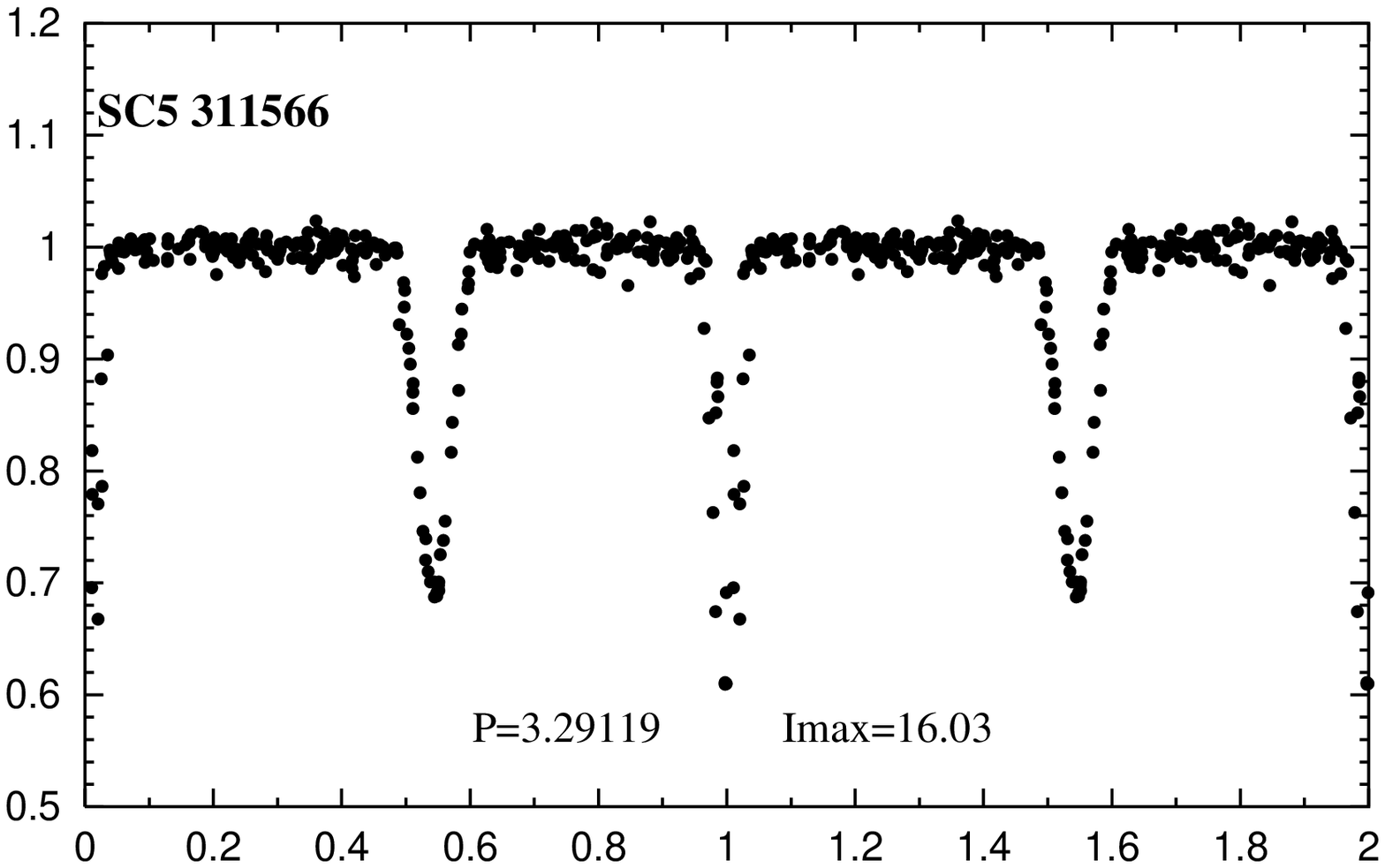}}
  \resizebox{\linewidth}{!}{\includegraphics{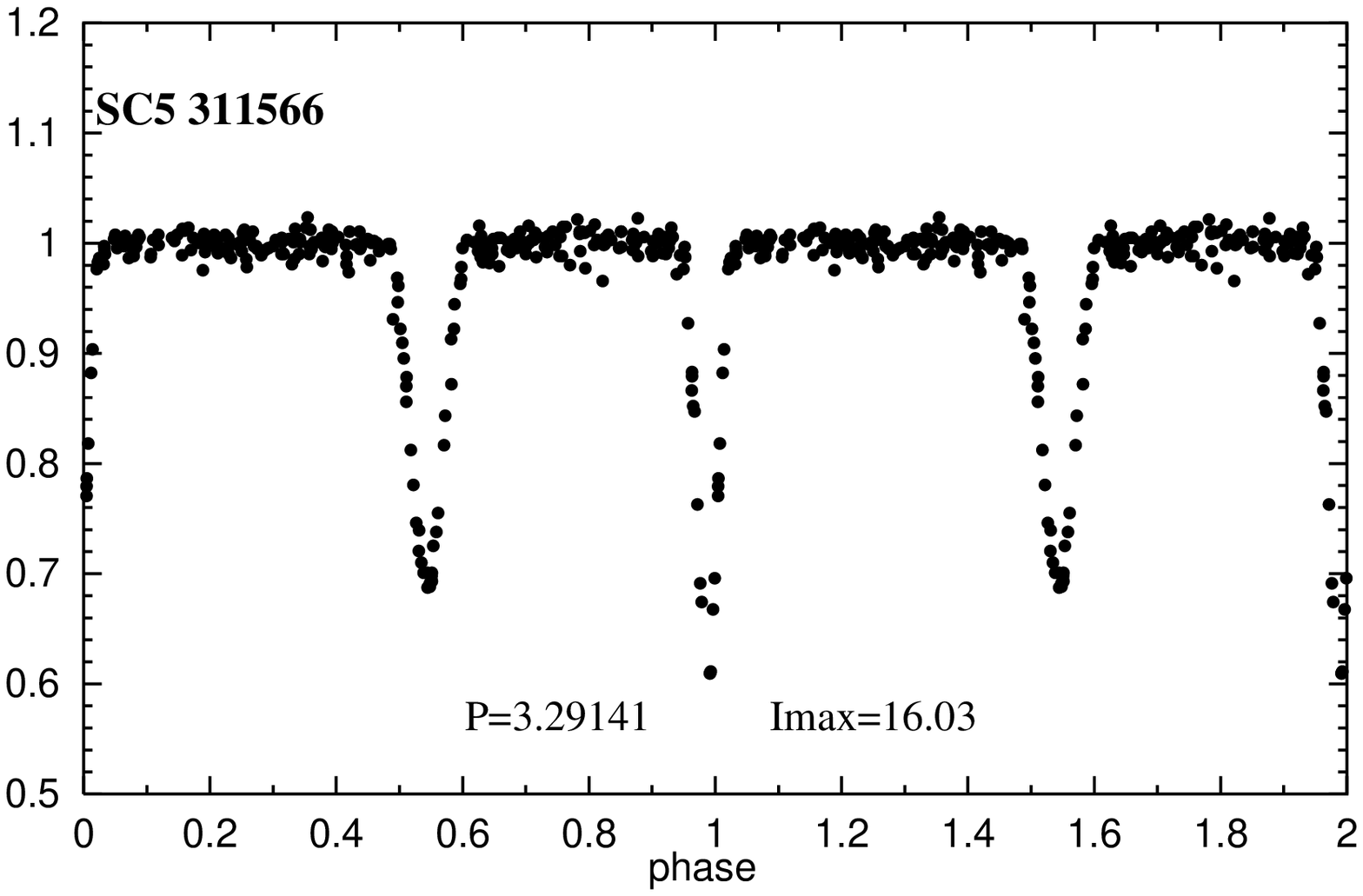}}
\end{minipage}
\caption{The light curves of SC5 311566 tuned to the shape of primary minimum 
(upper panel) and to the secondary minimum (middle panel). Lower panel: 
the light curve computed assuming a progressive shift of the secondary
minimum in respect to the primary minimum - see the text for explanation.}
\label{fig:311566com}
\end{figure}

%The conclusion is that SC1 25589 can be suggested as a one of the most 
%suitable distance indicators to SMC among the 17-th magnitude binaries.   

The candidates for the analysis were chosen after the visual inspection of 
the light curves of eclipsing binaries in SMC released by Udalski et al.~(1998).
A preliminary set of about 25 binaries were selected brighter than $I = 16.4$ mag at 
maximum brightness and with the depth of the primary and secondary 
eclipses greater than 0.25 mag. The candidates fall into three separate groups. 
The first group, hereafter called A, consist of well-detached, eccentric 
systems with, usually, deep narrow eclipses. The second group, hereafter 
called B, comprise the short-period systems with apparently circular orbits 
and the small separation between components. As a result, the systems from group B show quite
large proximity effects in their light curves. The third group C comprises 
well separated long period systems with circular orbits. As we expect long period 
systems to be, in general, eccentric this is most probably a chance selection.

\begin{table*}
\begin{center}
\begin{minipage}{\linewidth}
\caption{The selected eclipsing binaries}\label{tab:candidates}
\tabcolsep9pt
\begin{tabular}{@{}|l|c|c|c|c|c|c|c|@{}}
\hline
\hline
Object & $\alpha_{2000}$ & $\delta_{2000}$ & Period\footnote{Calculated by the period finding program. See $\S$\ref{period}.} & Epoch\footnote{Calculated from light curve analysis. See $\S$ 3.1}& $I$ & $B-V$ & $V-I$\\
 & (h:mm:ss) & (deg:mm:ss) & (days) & ($+2\, 450\, 000$) & & & \\\hline
\multicolumn{8}{c}{The A subset}\\
SC3 139376 & 0:44:08.69& $-$73:14:17.9 &  6.05213 &629.07158 &14.482& $-$0.165& $-$0.126 \\
SC5 129441 & 0:49:40.56& $-$73:00:22.5 &  8.05104 &474.20881 &16.012& $-$0.105& $-$0.072 \\
SC5 311566 & 0:51:34.84& $-$72:45:45.9 &  3.29132 &468.56556 &16.032& $-$0.120& $-$0.062 \\
SC6  11143 & 0:51:39.69& $-$73:18:44.8 &  5.72606 &471.95035 &16.251& $-$0.138& $-$0.092 \\
SC6  67221 & 0:52:06.22& $-$72:45:14.3 & $\!\!\!10.97765$ &466.88188 &15.327& $-$0.101& $-$0.095 \\
SC6 221543 & 0:53:40.39& $-$72:52:22.0 &  3.41680 &468.59872 &16.089& $-$0.218& $-$0.109 \\ 
SC6 272453 & 0:54:33.22& $-$73:10:39.6 &  5.70970 &469.30090 &15.805& $-$0.172& $-$0.057 \\
SC9 163575 & 1:02:46.27& $-$72:24:44.2 &  1.97109 &629.49776 &15.181& $-$0.248& $-$0.221 \\
SC10 37223 & 1:03:41.42& $-$72:03:06.9 &  2.58124 &630.63675 &16.336& $-$0.197& $-$0.164 \\
\multicolumn{8}{c}{The B subset}\\
SC4  53898 & 0:46:32.79 &$-$73:26:39.4&   1.74107 &624.86290 &16.201 &$-$0.128& $-$0.129 \\
SC4 103706 & 0:47:25.50 &$-$73:27:16.7&   1.35585 &623.94692 &15.416 &$-$0.195& $-$0.208 \\
SC4 163552 & 0:47:53.24 &$-$73:15:56.5&   1.54581 &625.35629 &15.774 &$-$0.046& $-$0.006 \\
SC5  38089 & 0:49:01.85 &$-$73:06:06.9&   2.38941 &469.77239 &15.261 &$-$0.238& $-$0.111 \\
SC6 215965 & 0:53:33.35 &$-$72:56:24.1&   3.94603 &473.26734 &14.133 &$-$0.234& $-$0.190 \\
SC6 319966 & 0:54:23.49 &$-$72:37:23.0&   1.06466 &466.68981 &16.363 &$-$0.151& $-$0.105 \\
SC8 104222 & 0:58:25.08 &$-$72:19:09.8&   1.56972 &629.30952 &16.042 &$-$0.144& $-$0.146 \\
SC9 175336 & 1:02:53.40 &$-$72:06:43.2&   2.98654 &628.36738 &14.867 &$-$0.222& $-$0.192 \\ 
\multicolumn{8}{c}{The C subset}\\
SC4 192903 & 0:48:22.69 &$-$72:48:48.6& $\!\!\!\!\!\!102.86800$ &761.00300 &16.242 & $\,$0.771&  $\,$0.956 \\
SC8 201484 & 1:00:18.05 &$-$72:24:07.5& $\!\!\!\!\!\!185.19000$ &641.56700 &14.230 & $\,$0.883&  $\,$0.977 \\\hline
\end{tabular}
\end{minipage}
\end{center}
\end{table*}

Afterwards, the photometric data for each selected star were extracted, 
using the coordinates as identifier. Some stars were not found 
in OGLE-DIA database because they probably passed undetected through the variability 
checking filters (I.~Soszy{\'n}ski 2002, private communication).
The Table~\ref{tab:candidates} 
gives the names and identification of 19 selected systems together with a summary of photometric 
parameters.  

\subsection{Initial data preparation} 
%-----------------------------------------
\label{period}The observations were folded with trial periods following the method given by 
Kaluzny et al.~(1998). The searches were usually restricted to a period close 
to the period given by Udalski et al.~(1998). The final period was found by tuning to the
shape of the minima. It was found that in most cases the resulting 
periods agreed very well with those given by Udalski et al.~(1998). However, 
for SC6 67221, I have found a new period resulting from observations of both 
eclipses. This binary turned out to be a very eccentric system with the orbital 
period of 10.977 days. The period is close to a whole number of 11 days and 
this is the reason for spurious secondary eclipse detection in the original OGLE 
catalog. Fig.~\ref{fig:67221com} presents the light curve based on the proper 
period and, for comparison, the light curve from the original OGLE catalog.

Two eccentric binaries SC3 139376 and SC5 311566 show small changes  
in the light curve most probably due to the apsidal motion. 
In both cases the tuning to the shape of the primary and the secondary minimum
gives significantly different periods -- see as an example SC5 311566 (Fig.\ref{fig:311566com}).   
Smooth light curves were obtained assuming that the change of the shape of
minima is small and the secondary minimum was progressively shifting in respect to the
primary minimum during the period of observations. The linear correction was applied 
to all phases corresponding to the secondary eclipse (phases: 0.35-0.55 for SC3 139376 and 
0.4-0.6 for SC5 311566) in the form:
\begin{equation}
\phi_i = \phi_i + \alpha \times E, \label{shift}
\end{equation}
where $\phi$ and $E$ are the phase and the epoch number of the $i$-th
observation. Denoting the period resulting from the tuning to the primary eclipse by $P_p$ 
and to the secondary one by $P_s$, the factor $\alpha$ was calculated by formula: $\alpha=(P_p- 
P_s)/P_p$ and was $7.5\cdot 10^{-5}$ and $7.0\cdot 10^{-5}$ for
SC3 139376 and SC5 311566, respectively.

The old reference 
epochs $T_o$ given by Udalski et al.~(1998) were used to fold up the observations. 
New reference epochs were 
determined later during the light curve analysis -- see $\S$ 3.1. For each star I 
calculated out-of-eclipse $I$ magnitude at the maximum of the brightness. 
The DIA data does not include the $B$ and $V$ photometry. For the purpose of 
obtaining the most accurate $B-V$ and $V-I$ colours I recalculated the out-of-eclipse 
$B$, $V$ and $I$ magnitudes using the old DoPHOT photometry. Note, that the magnitudes 
which I have found refer to the mean magnitudes at the maximum 
brightness rather than to the maximal magnitudes given in the 
original OGLE catalog. Resulting out-of-eclipse $I$ magnitudes were then used to 
normalize the light curves.        

\begin{figure*}
\begin{minipage}[ht]{0.46\linewidth}
   \hspace*{0.9cm}\resizebox{\linewidth}{!}{\includegraphics{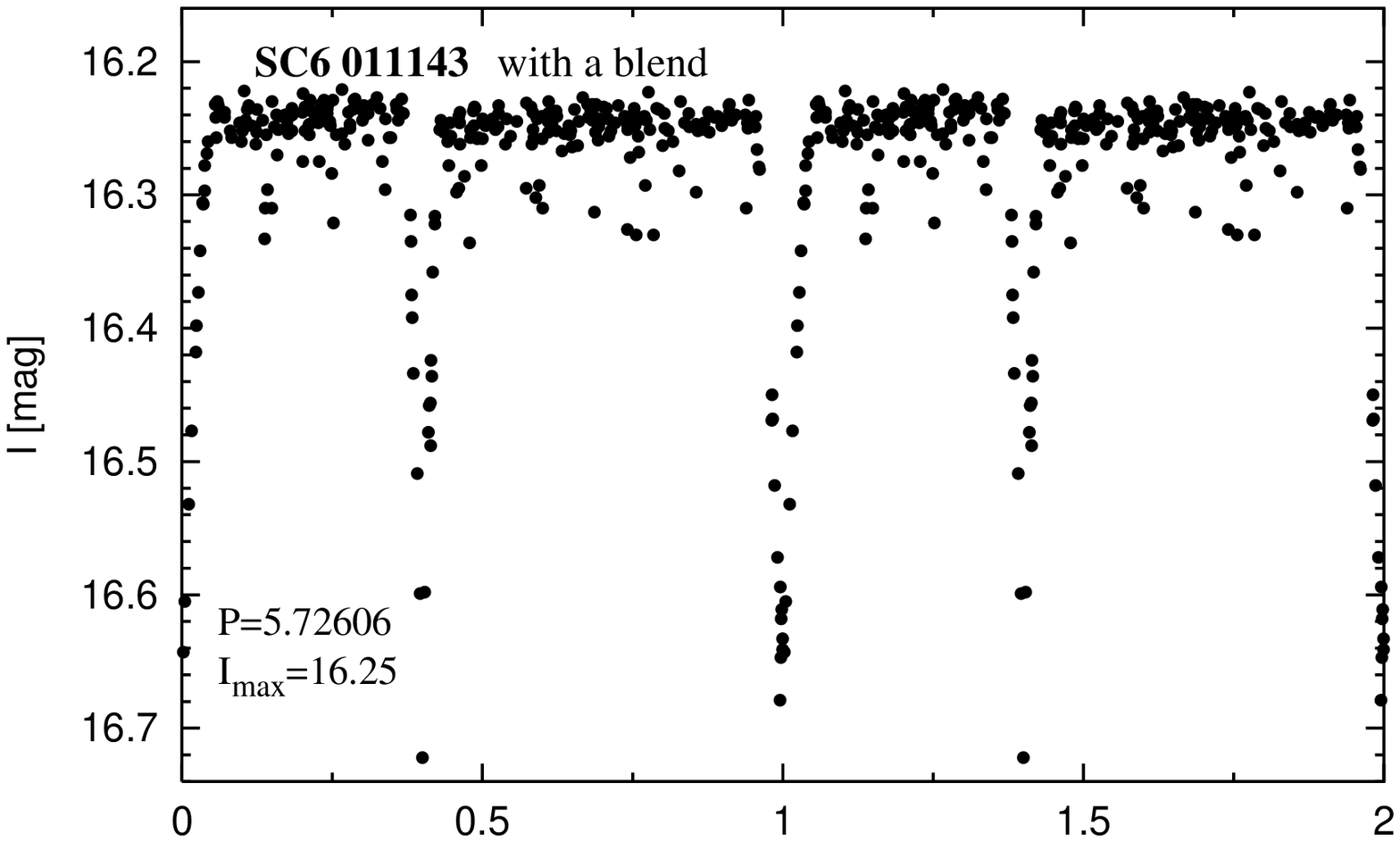}}
\mbox{}\\[-0.6cm]
  \hspace*{1cm}\resizebox{0.99\linewidth}{!}{\includegraphics{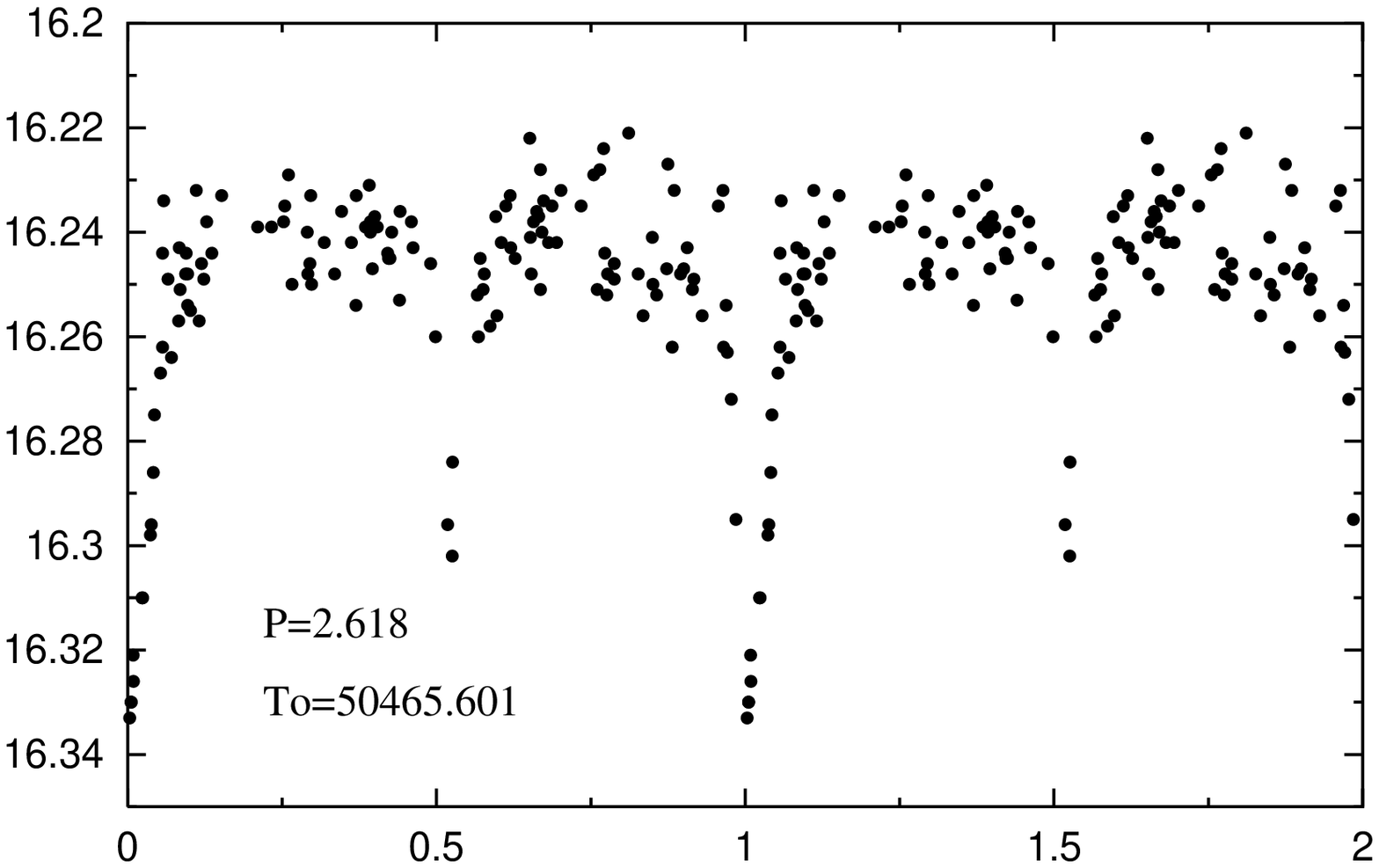}}
\mbox{}\\[-0.6cm]
   \hspace*{0.9cm}\resizebox{\linewidth}{!}{\includegraphics{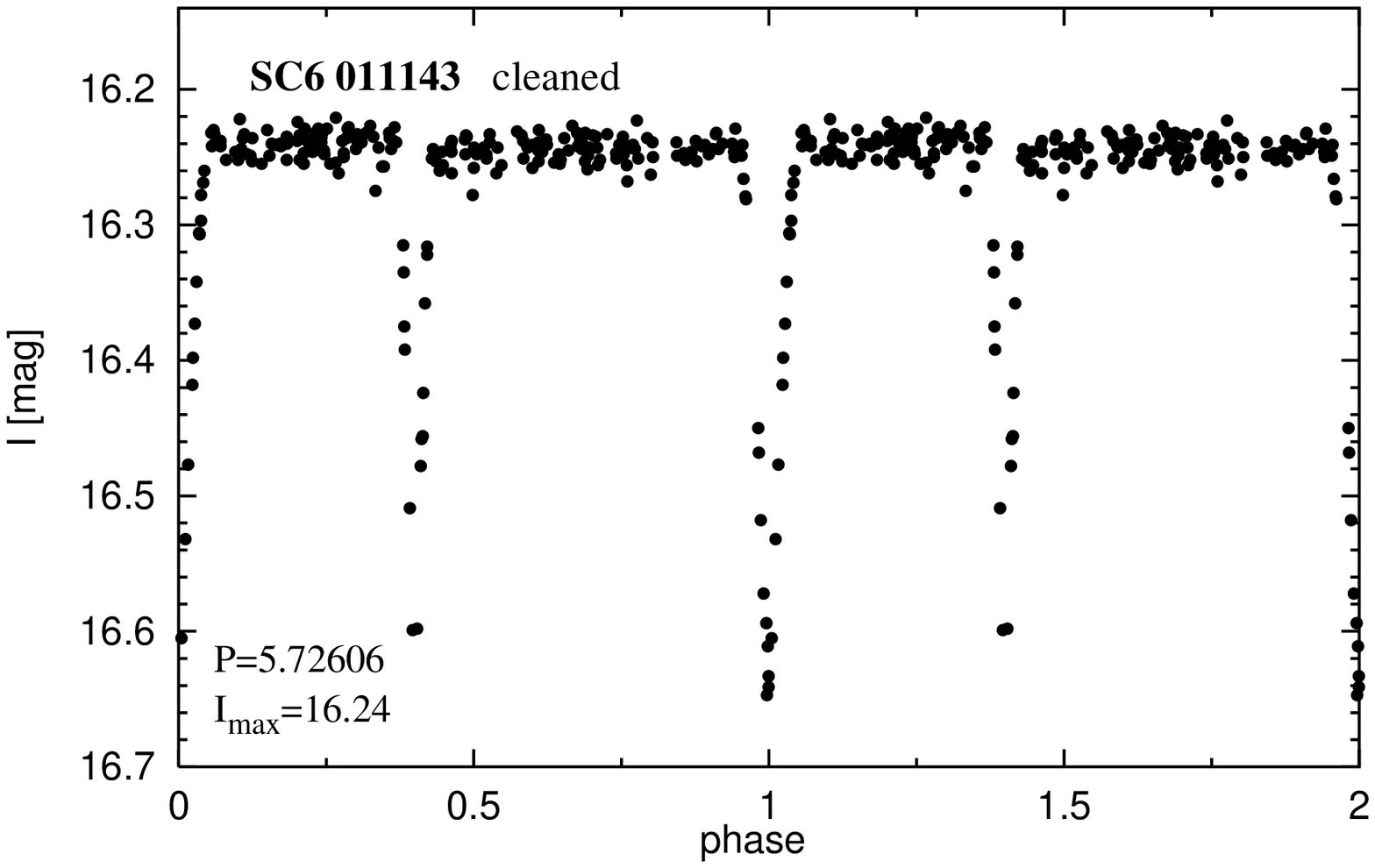}}
\end{minipage}\hfill
\begin{minipage}[h]{0.46\linewidth}
   \hspace*{-0.9cm}\resizebox{\linewidth}{!}{\includegraphics{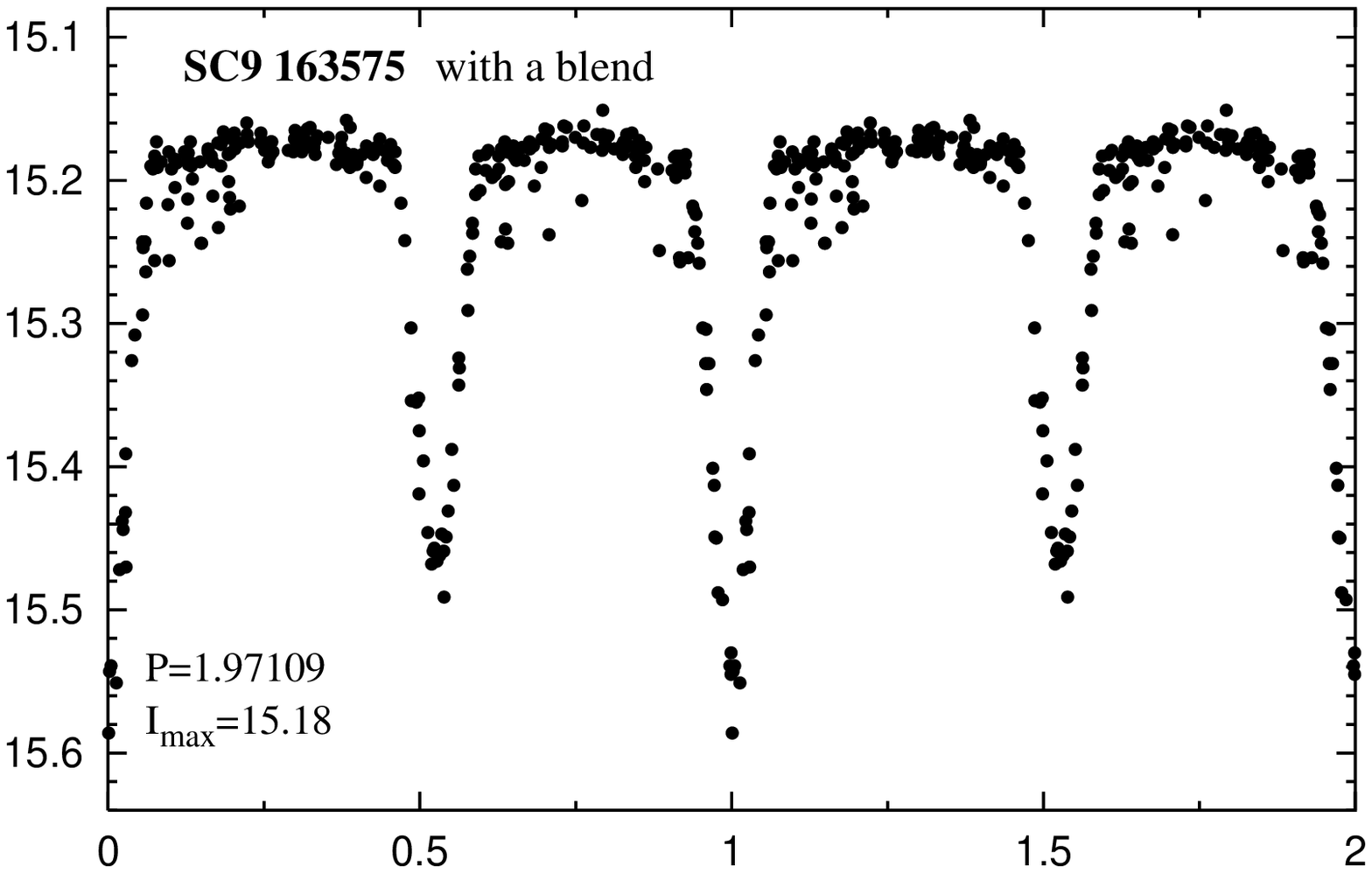}}
\mbox{}\\[-0.24cm]
 \hspace*{-0.8cm}\resizebox{0.99\linewidth}{!}{\includegraphics{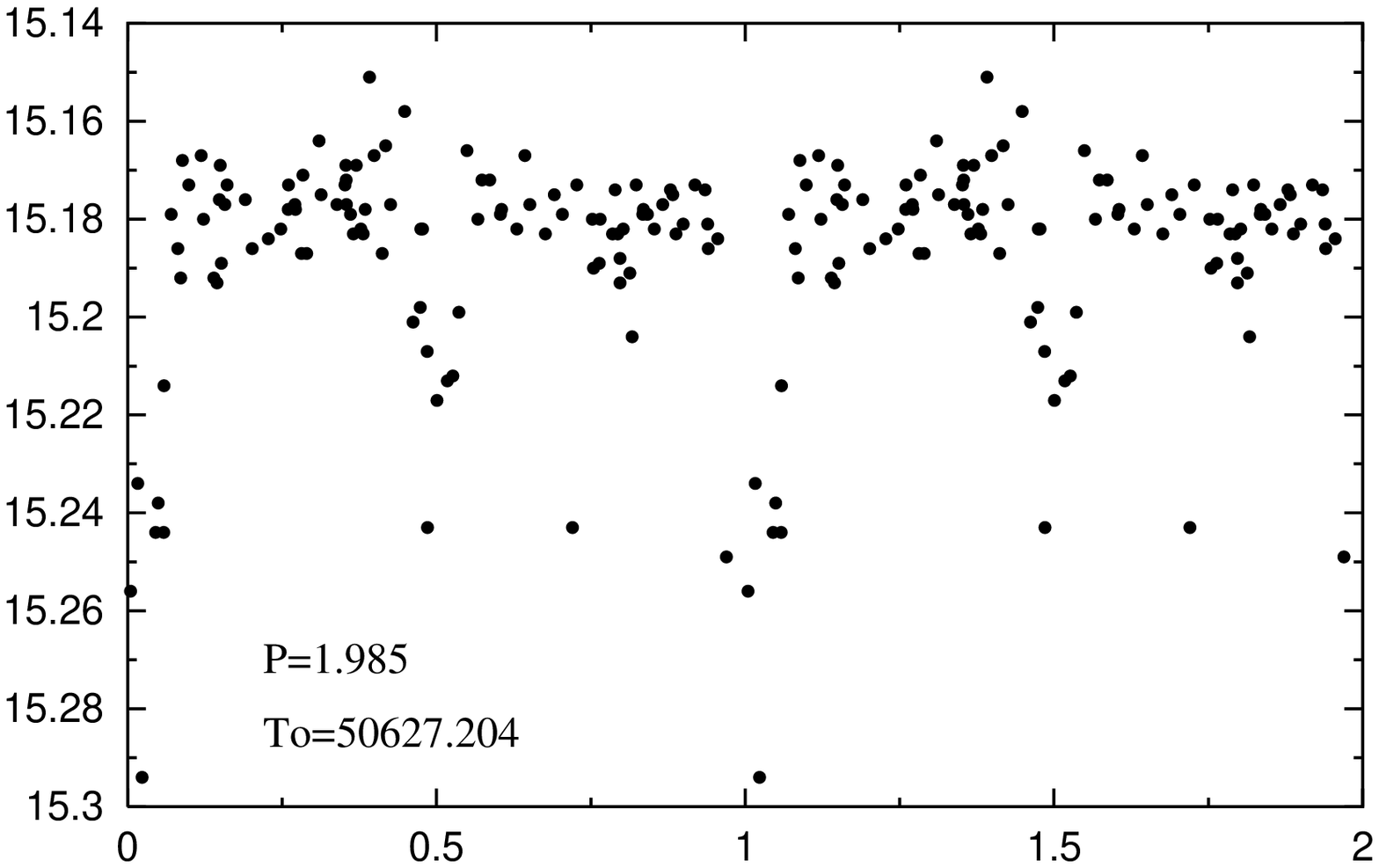}}
\mbox{}\\[-0.24cm]
   \hspace*{-0.9cm}\resizebox{\linewidth}{!}{\includegraphics{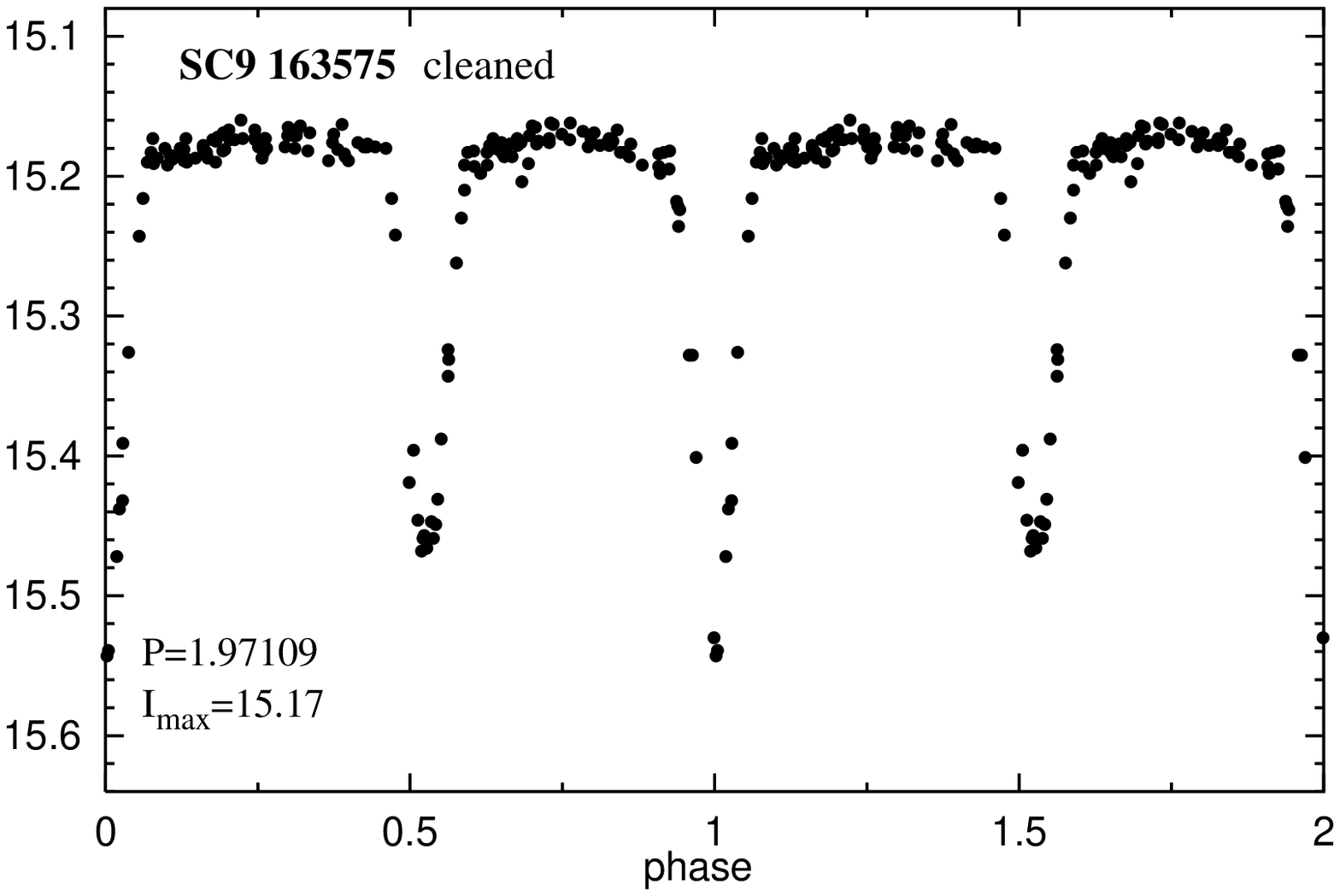}}
\end{minipage}
\caption{Upper panel: the original DIA light curves
of SC6 11143 and SC9 163575. Middle panel: the out-of-eclipse observations
of both stars folded up with a period and an epoch given in each picture.   
Both light curves suggest that a blend is an eclipsing binary. 
Lower panel: the final light curves after removing the variability
of the blend.}
\label{fig:blends}
\end{figure*}

In two cases (SC6 11143 and SC9 163575) the 
light curve shows considerable amount of scatter, much larger than the average
error of the observations. The light curves of both systems are presented on 
Fig.~\ref{fig:blends} (upper panel). I have removed observational points when the eclipses 
occurred and rerun the period finding program to find some periodicity in the 
scatter. It turns out that in the case of SC6 11143 the out-of-eclipse observations
assemble with the period of 2.618 days and form a light curve
with two well-defined, narrow minima of different depth (middle panel of Fig.~\ref{fig:blends}). 
It is likely that the binary SC6 11143  blends with another short period and slightly
eccentric eclipsing binary. The similar analysis done for the system SC9 163575 shows that the 
out-of-eclipse observations assemble with the period of 1.985 days and form 
two minima separated by a half of the period. It seems that the system 
SC9 163575 blends also with a short period eclipsing binary. Their periods
are very close to each other and in the original light curve the points,
corresponding to the minima of the blend, form two clumps situated
around the minima of SC9 163575. The light curves of
both systems were cleaned by removing from the light curve all points corresponding
to the moments when the eclipses occurred in the blended binary. The resulting light
curves are presented in the lower panel of Fig.~\ref{fig:blends}.
Also the system SC5 38089 in spite of its relatively high brightness 
($I_{max}=15.26$) has quite large scatter of observations. The analysis done 
for this system shows that the changes of the brightness are randomly distributed
and thus, it is likely that the reason of the scatter is some kind of an 
intrinsic variability.

\section{Light curve analysis}
%-----------------------------------
\subsection{Method}
%------------------------
All fits to the light curves were obtained using a modified version of the 
Wilson-Devinney(1971, Wilson 1979, 1992, hereafter WD) program with a model atmosphere 
procedure based on the ATLAS9 code and developed by Milone, Stagg \& Kurucz (1992).
An automated iterations scheme of WD program was employed. For A subset a detached 
configuration (mode 2) was chosen for all solutions and simple reflection 
treatment was used (MREF=1, NREF=1). In the case of B subset both detached and semi-detached 
(mode 5) configurations were used to obtain solutions. For B and C subsets the detailed 
reflection treatment with a double reflection was taken into account (MREF=2, NREF=2).
For all stars $\log g = 4.0$ was fixed
except for stars from C subset for which $\log g = 2.0$ was assumed.

I have found the initial guess via trial-and-error fitting of the light curves. 
The convergence is defined to have been achieved if the parameters corrections 
given by WD program are smaller than 0.75 times the standard errors (1.0 in the 
case of eccentricity and argument of periastron) on two consecutive iterations. 
The adjusted parameters were the surface potentials ($\Omega_1,\, \Omega_2$), 
the effective temperature of the secondary ($T_2$), the luminosity of the primary 
($L_1$), orbital inclination ($i$) and the zero epoch offset ($\Delta\phi$). 
For A subset I have adjusted two more parameters: eccentricity ($e$) and the 
argument of periastron ($\omega_o$).

If for an eccentric system the convergence was not achieved I suppressed the adjustment
of $e$, $\omega$ and $\Delta\phi$ provided the 
correction were smaller than the standard error for these parameters 
on three consecutive iterations. It turned out that such procedure allowed for 
better convergence and the solution was achieved in all cases. 
Afterwards I have run additional iterations, allowing  all parameters to vary, 
to find error estimates of the final solution. Similar procedure was
employed for the systems from 
B and C subsets, just fixing the circular orbit and the suppressing included only the zero phase offset.               
Among the systems in B subset the third light ($l_3$) adjustment was needed 
to get the convergence in most cases. For all systems correction $\Delta T_o$ to the epoch
given by Udalski et al.~(1998) was applied in the form: \( \Delta T_o = \Delta\phi \cdot P \), 
where $\Delta\phi$ is the zero epoch offset of the final solution and $P$ is the period of
the system. 

WW1 introduced a quantity $F_e$ which is greater than unity for systems with
complete eclipses:
\begin{equation}
F_e=\frac{r_g + r_s - \cos{i}}{2 r_s}, \label{fe}
\end{equation}
where $r_g$ is the fractional the radius of larger star and $r_s$ is the
fractional radius of the smaller star, both expressed in units of semi-major
axis $A$. 
$F_e$ is valid, in the form presented above,
for circular orbits or perpendicular orbits ($i$=90$^\circ$) and also 
for eccentric orbits provided the
fractional radii of components are expressed in units of the distance 
between stars {\it at the moment} of the mid-eclipse.  
In general case of eccentric orbits, if one eclipse occurs in the neighbourhood of
periastron and is complete, the other may be partial and
in the extreme cases, there may be an eclipse near periastron and none near
apastron.  
The instantaneous distance between components $d$, in units of semi-major 
axis $A$, is given by the equation:
\begin{equation}
d=\frac{1-e^2}{1+e\cos{\nu}}\, , \label{distance}
\end{equation}
where $\nu$ is the true anomaly.
The primary and secondary minima occur when the true 
anomalies $\nu_1$ and $\nu_2$ are
(neglecting higher powers of $e$, see Kopal 1959, Chapter 6, equation (9-22)):
\begin{eqnarray}
\nu_1 &=& 90^\circ -\, \omega -\, e\cos{\omega} \cot^2{i}\, (1-e\sin{\omega} \csc^2{i}),  \\
\nu_2 &=& 270^\circ -\, \omega +\, e\cos{\omega} \cot^2{i}\, (1-e\sin{\omega} \csc^2{i}), 
\end{eqnarray}
respectively. Substituting $\nu_1$ and $\nu_2$ into Eq.~\ref{distance} the distances 
were calculated at the moments of both mid-eclipses. Then appropriate
quantities $F^1_e$ and $F^2_e$ referring to the primary and the secondary
eclipse, respectively, were computed.  

Semidetached configuration (mode 5) were investigated for all stars from the B subset 
in order to check which systems are definitively separated. When the semidetached 
configuration was investigated the mass ratio ($q$) of the components was
adjusted and secondary surface potential ($\Omega_2$) was set by $q$.
However, it turned out that no solution could be found in this
mode for any star from B subset in spite of many tests for different
sets of adjustable parameters. Accordingly, in subsequent part
of the paper I will refer only to detached solutions found in mode 2.

\begin{table*}
\begin{minipage}{\linewidth}
\caption{Photometric parameters}\label{tab:solA}
\tabcolsep3pt
\begin{tabular}{@{}|l|c|c|c|c|c|c|c|c|c|c|@{}}
\hline
\hline
     &Radius &Radius & Temperature & Inclination& Eccentricity & Argument of & Third & Luminos. & \\
Object & $R_1/A$ & $R_2/A$ & ratio $T_2/T_1$ & $i$ &$e$ & periastron $\omega$&light $l_3$ & ratio $l_2/l_1$
&$F^1_e$ & $F^2_e$ \\\hline
\multicolumn{11}{c}{The A subset}\\
SC3  139376$^\ast$ & 0.177$\pm$0.004& 0.194$\pm$0.004& 1.007$\pm$0.006& 83.3$\pm$0.2& 0.064$\pm$0.003& 133.7$\pm$5.7&
0.0&1.191& 			 0.734 & 0.704  \\
SC5  129441$^\ast$$^\dagger$ & 0.158$\pm$0.001& 0.162$\pm$0.002&0.999$\pm$0.001& 85.2$\pm$1.0&0.375$\pm$0.001&172.0$\pm$0.4& $\,\;$0.0$^\ddag$& 1.079&   0.797 & 0.773  \\
SC5  311566 & 0.157$\pm$0.004& 0.147$\pm$0.002& 0.962$\pm$0.005& 85.7$\pm$0.6& 0.251$\pm$0.004& 070.0$\pm$0.1& 0.0& 0.843&				 0.841 & 0.721  \\ 
SC6   11143$^\dagger$ & 0.133$\pm$0.004& 0.112$\pm$0.002& 0.972$\pm$0.020& 88.8$\pm$0.1& 0.263$\pm$0.001& 234.3$\pm$0.6& 0.30$\pm$0.05& 0.704&		 0.983 & 1.022  \\
SC6   67221$^\ast$ & 0.131$\pm$0.002& 0.119$\pm$0.002&1.154$\pm$0.011& 83.9$\pm$0.3&0.401$\pm$0.002&043.5$\pm$0.9& $\,\;$0.0$^\ddag$& 1.058&		 0.757 & 0.533  \\
SC6  221543 & 0.202$\pm$0.006& 0.182$\pm$0.005& 0.960$\pm$0.007& 84.2$\pm$0.7& 0.099$\pm$0.002& 146.1$\pm$2.7& 0.0& 0.745&				 0.794 & 0.764  \\
SC6  272453 & 0.149$\pm$0.004& 0.180$\pm$0.004& 0.961$\pm$0.006& 85.2$\pm$0.6& 0.035$\pm$0.004& 236.0$\pm$4.5& 0.0& 1.392&				 0.815 & 0.831  \\
SC9  163575 & 0.234$\pm$0.004& 0.225$\pm$0.004& 0.891$\pm$0.007& 80.6$\pm$0.4& 0.043$\pm$0.002& 010.7$\pm$7.4& 0.10$\pm$0.05& 0.762&			 0.661 & 0.655  \\
SC10  37223 & 0.202$\pm$0.004& 0.195$\pm$0.004& 0.951$\pm$0.006& 83.8$\pm$0.3& 0.087$\pm$0.003& 142.3$\pm$2.4& 0.0& 0.874&				 0.757 & 0.728  \\
\multicolumn{11}{c}{The B subset}\\
SC4   53898$^\ast$ & 0.189$\pm$0.001& 0.356$\pm$0.002& 0.910$\pm$0.006& 85.7$\pm$1.2 & 0.0& 090.0& 0.187$\pm$0.012& 3.035 & \multicolumn{2}{c}{1.210} \\
SC4  103706$^\dagger$ & 0.350$\pm$0.002& 0.284$\pm$0.006& 0.964$\pm$0.004& 86.8$\pm$2.0& 0.0& 090.0& 0.382$\pm$0.014& 0.612 & \multicolumn{2}{c}{1.003}  \\ 
SC4  163552$^\dagger$ & 0.338$\pm$0.006& 0.326$\pm$0.008& 1.004$\pm$0.008& 85.4$\pm$0.9& 0.0& 090.0& 0.263$\pm$0.007& 0.951 & \multicolumn{2}{c}{0.888}  \\  
%SC5   38089 & 0.269$\pm$0.006& 0.224$\pm$0.009& 0.985$\pm$0.003& 78.7$\pm$2.4 & 0.0& 090.0& 0.108$\pm$0.067& 0.654 & \multicolumn{2}{c}{0.653}  \\ 
SC5 38089   & 0.246$\pm$0.009& 0.255$\pm$0.007& 0.988$\pm$0.003& 76.8$\pm$1.9 & 0.0& 090.0& 0.0 & 1.053 & \multicolumn{2}{c}{0.545}  \\
SC6  215965$^\ast$$^\dagger$ & 0.277$\pm$0.003& 0.335$\pm$0.002&0.922$\pm$0.001& 75.6$\pm$0.3&0.0&090.0& $\,\;$0.0$^\ddag$ & 1.356 & \multicolumn{2}{c}{0.656} \\
SC6  319966$^\dagger$ & 0.339$\pm$0.006& 0.324$\pm$0.008& 1.007$\pm$0.007& 88.3$\pm$3.0& 0.0& 090.0& 0.462$\pm$0.005& 0.861 & \multicolumn{2}{c}{0.973}  \\   
SC8  104222 & 0.311$\pm$0.011& 0.331$\pm$0.009& 0.993$\pm$0.008& 86.8$\pm$1.4& 0.0& 090.0& 0.087$\pm$0.013& 1.129 & \multicolumn{2}{c}{0.894}  \\   
SC9  175336$^\ast$ & 0.194$\pm$0.002& 0.358$\pm$0.002& 0.965$\pm$0.005& 88.7$\pm$0.8& 0.0& 090.0& 0.174$\pm$0.010& 3.228 & \multicolumn{2}{c}{1.330} \\
\multicolumn{11}{c}{The C subset}\\
SC4  192903 & 0.111$\pm$0.004& 0.145$\pm$0.003& 0.927$\pm$0.004& 87.2$\pm$0.7& 0.0 &090.0 &0.0& 1.165& \multicolumn{2}{c}{ 0.714}  \\
SC8  201484 & 0.171$\pm$0.003& 0.224$\pm$0.002& 0.888$\pm$0.002& 78.6$\pm$0.2& 0.0 &090.0 &0.0& 1.107& \multicolumn{2}{c}{ 0.569}  \\\hline    
\end{tabular}
\end{minipage}
\medskip
\begin{minipage}{\linewidth}
{\footnotesize The stars chosen in this paper as likely distance
indicators are marked by an asterisk. The stars indicated by Wyithe \&
Wilson (2001, 2002) are marked by a dagger. A double dagger means
that the binary has most probably no third light contribution.}
\end{minipage}
\end{table*}

\subsection{Parameter choice}
%-------------------------------
In the analysis presented here I have included the temperature of binary's 
components resulting from an intrinsic colour $(B-V)_o$ of the system. 
All $(B-V)$ colours from Table~\ref{tab:candidates} were dereddened using 
the average reddening toward SMC $E(B-V)=0.087$ (Massey et al.~1995, Udalski 2000). 
I set the temperature of the primary ($T_1$) respectively 
to the $(B-V)_o$ colour of the system using Flower's (1996) empirical calibration.  
Almost all the selected binaries have minima 
of similar depth what indicates that the difference between the temperatures 
of the components is small. Thus such a procedure is
justifiable. For those stars for which $(B-V)_0$ colour suggested a very high 
temperature ($\sim 4\, 10^{4}$ K for SC5 38089, SC6 215965, SC9 163575, SC9 175336) I set
up $T_1=35\, 000$ K.
  
I used a logarithmic law for the limb darkening (e.g.~van Hamme 1993). 
The $I$ limb darkenings $x_{1,2},\, y_{1,2}$ and bolometric limb darkenings 
$x^{bol}_{1,2},\, y^{bol}_{1,2}$  were calculated from 
van Hamme's (1993) tables of limb darkenings for appropriate $\log{g}$ and
$T$.
During the iterations the secondary's limb darkening coefficients were calculated
according to its temperature $T_2$.
The bolometric albedo and gravity brightening exponent were set to the canonical 
value of 1.0 for stars with radiative envelopes (von Zeipel 1924). However, 
for two red long-period systems SC4 192903 and SC8 201484 values typical for 
convective envelopes were assumed -- bolometric albedo was set to 0.5 
and gravity brightening exponent to 0.32. The synchronous rotation of both 
components was assumed ($F_1=F_2=1$). No level depending weighting of 
observations was applied -- parameter NOISE was set to 0.      

The mass ratio was assumed to be $q=m_2/m_1=1$ in all calculations. 
WW2 pointed out that in the case of detached 
eclipsing binary the light curve is almost insensitive to the mass ratio 
parameter. Only if one of the components is highly evolved (when its radius 
$R$ increase to about a half of the total separation $A$) or if the system 
has a very low mass ratio, the light curve becomes sensitive for this parameter. 
The preliminary tests show that no component from the set of binaries 
has fractional radius $r_1=R_1/a$ larger than 0.4. However, for two systems
from B subset it was necessary to adjust the mass ratio in order to get a good
solution (see \S 3.4).

\subsection{Solutions for A subset}
\begin{figure*}
\begin{minipage}[ht]{0.5\linewidth}
  \resizebox{\linewidth}{!}{\includegraphics{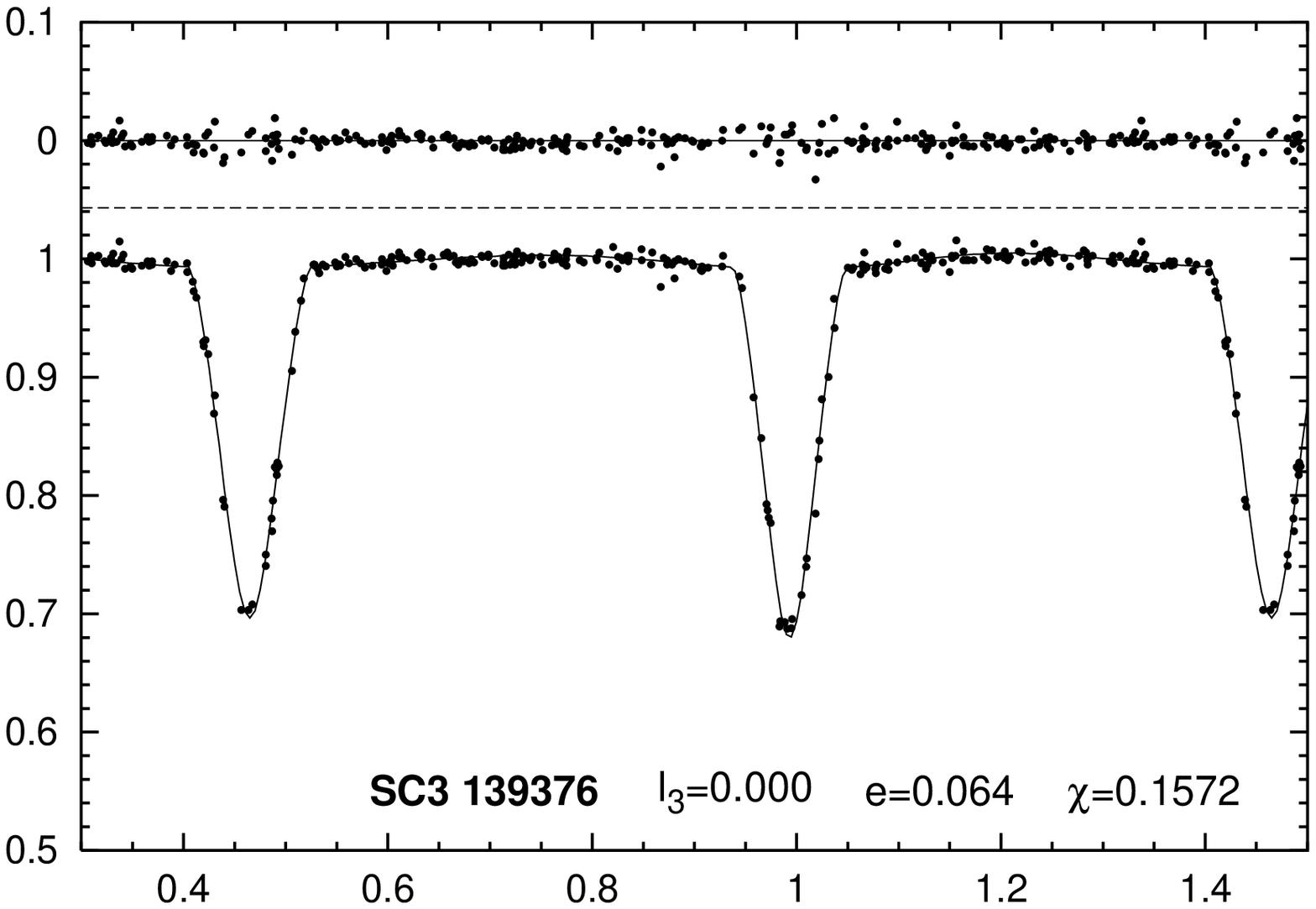}}
\mbox{}\\[-0.55cm]
  \resizebox{\linewidth}{!}{\includegraphics{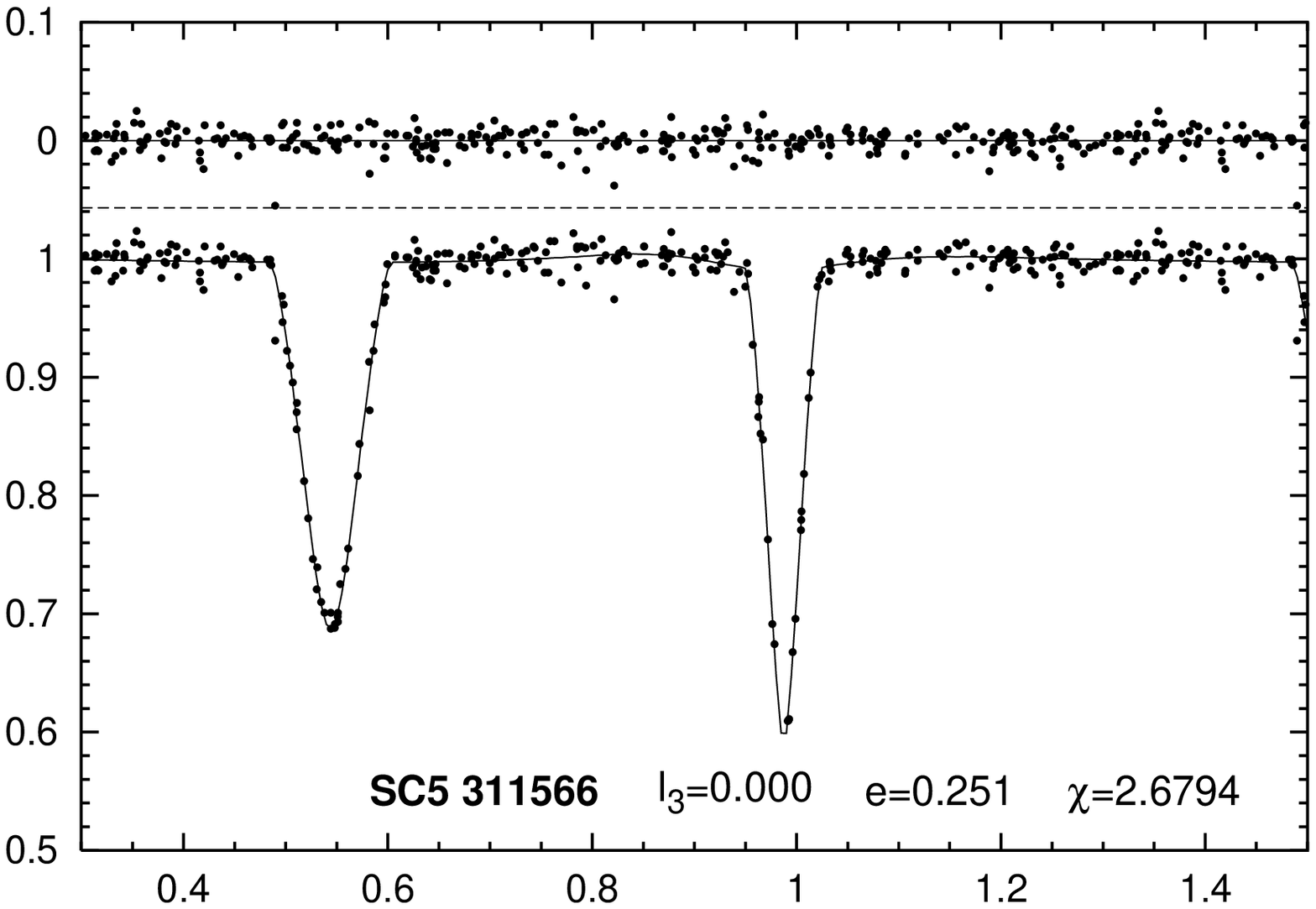}}
\mbox{}\\[-0.55cm]
  \resizebox{\linewidth}{!}{\includegraphics{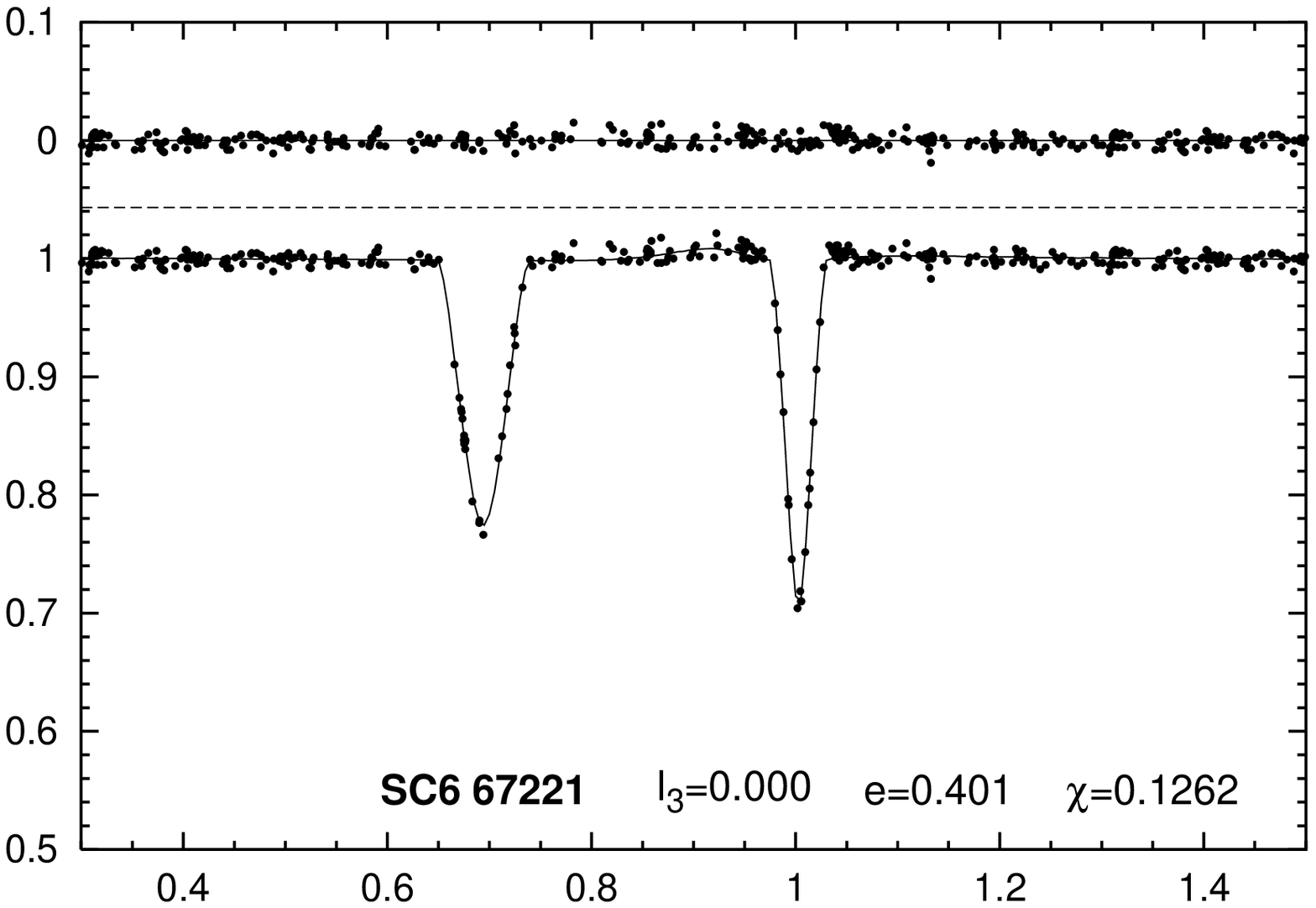}}
\mbox{}\\[-0.55cm]
  \resizebox{\linewidth}{!}{\includegraphics{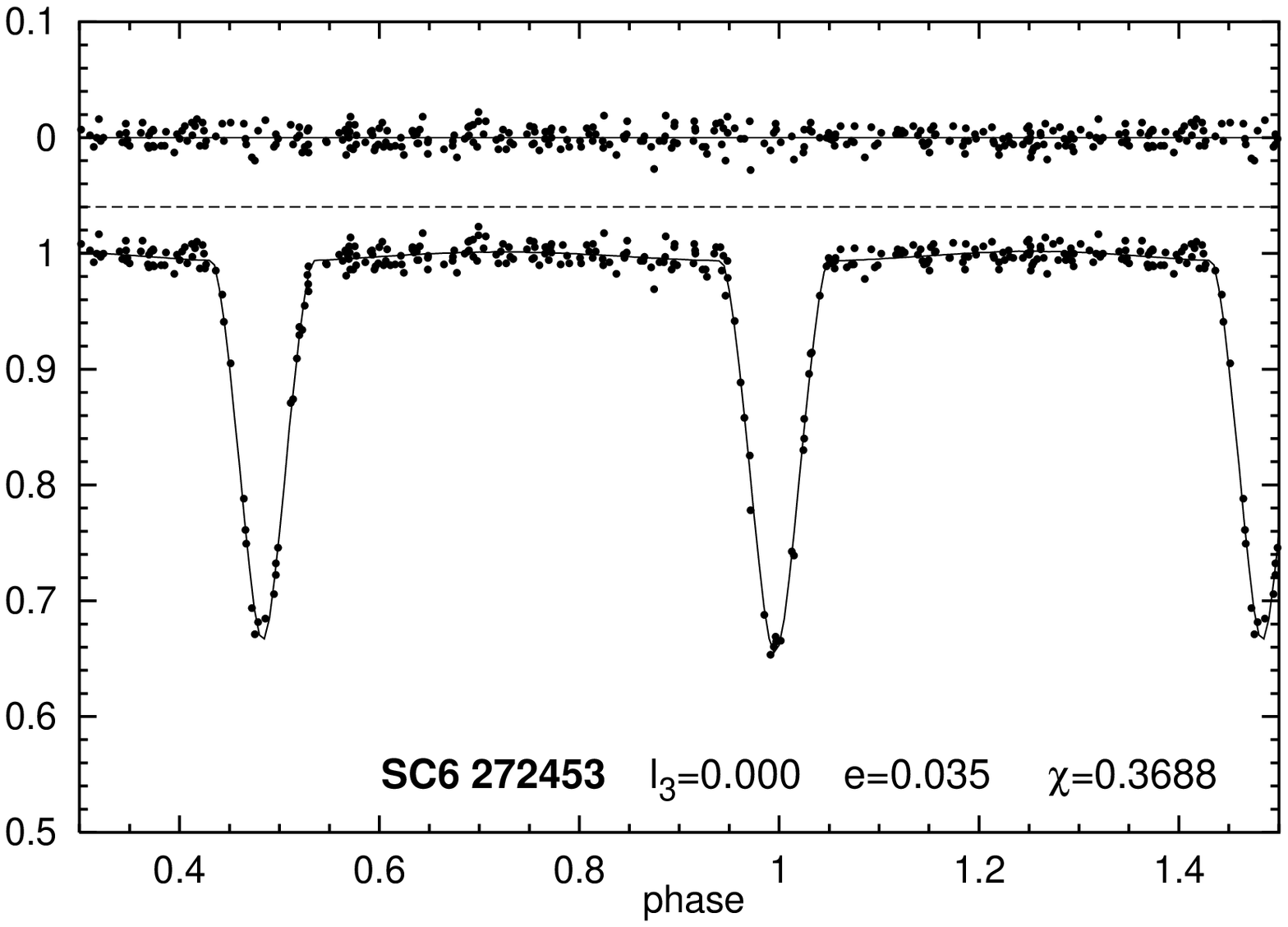}}
\end{minipage}\hfill
\begin{minipage}[h]{0.5\linewidth}
  \resizebox{\linewidth}{!}{\includegraphics{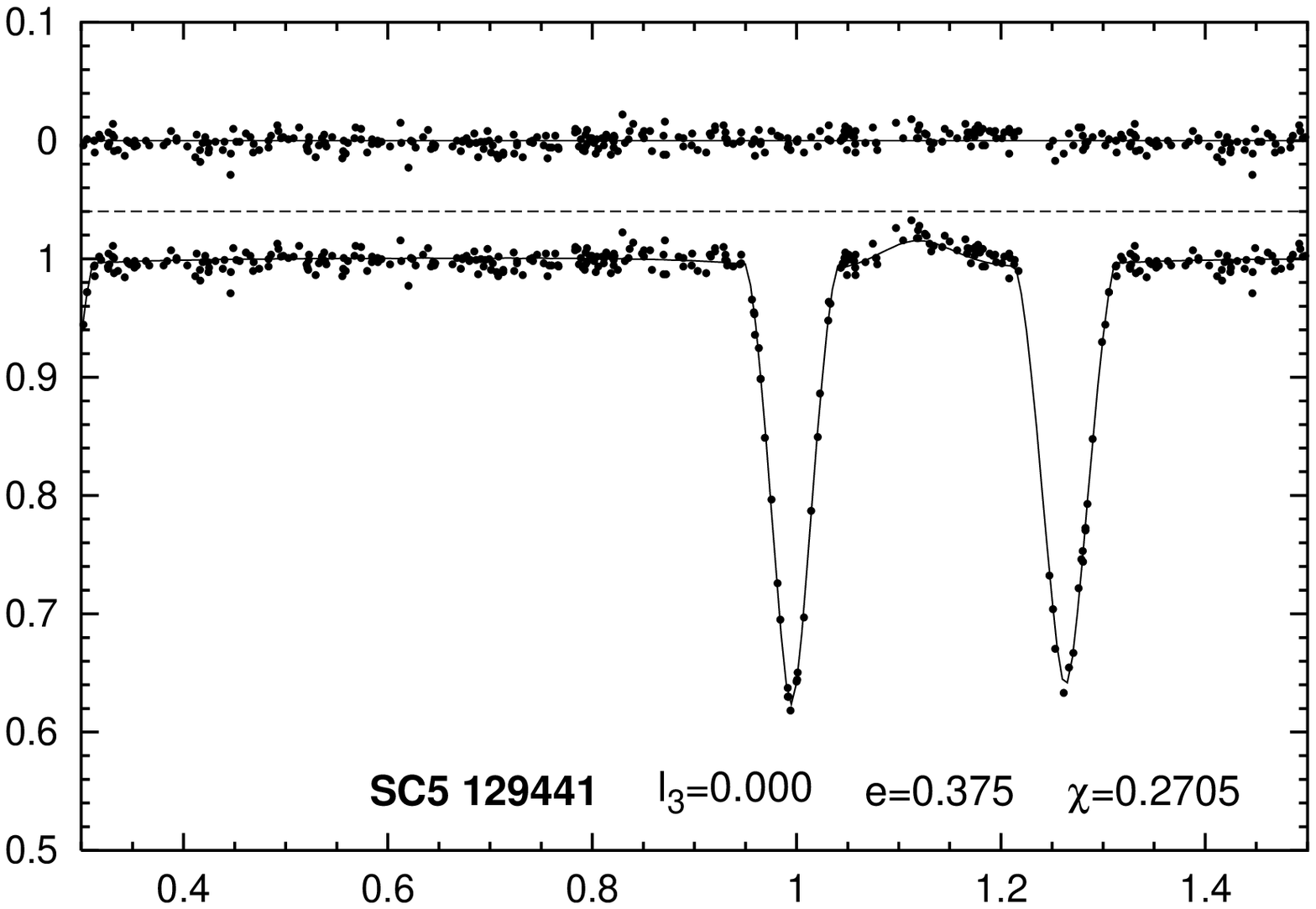}}
\mbox{}\\[-0.55cm]
  \resizebox{\linewidth}{!}{\includegraphics{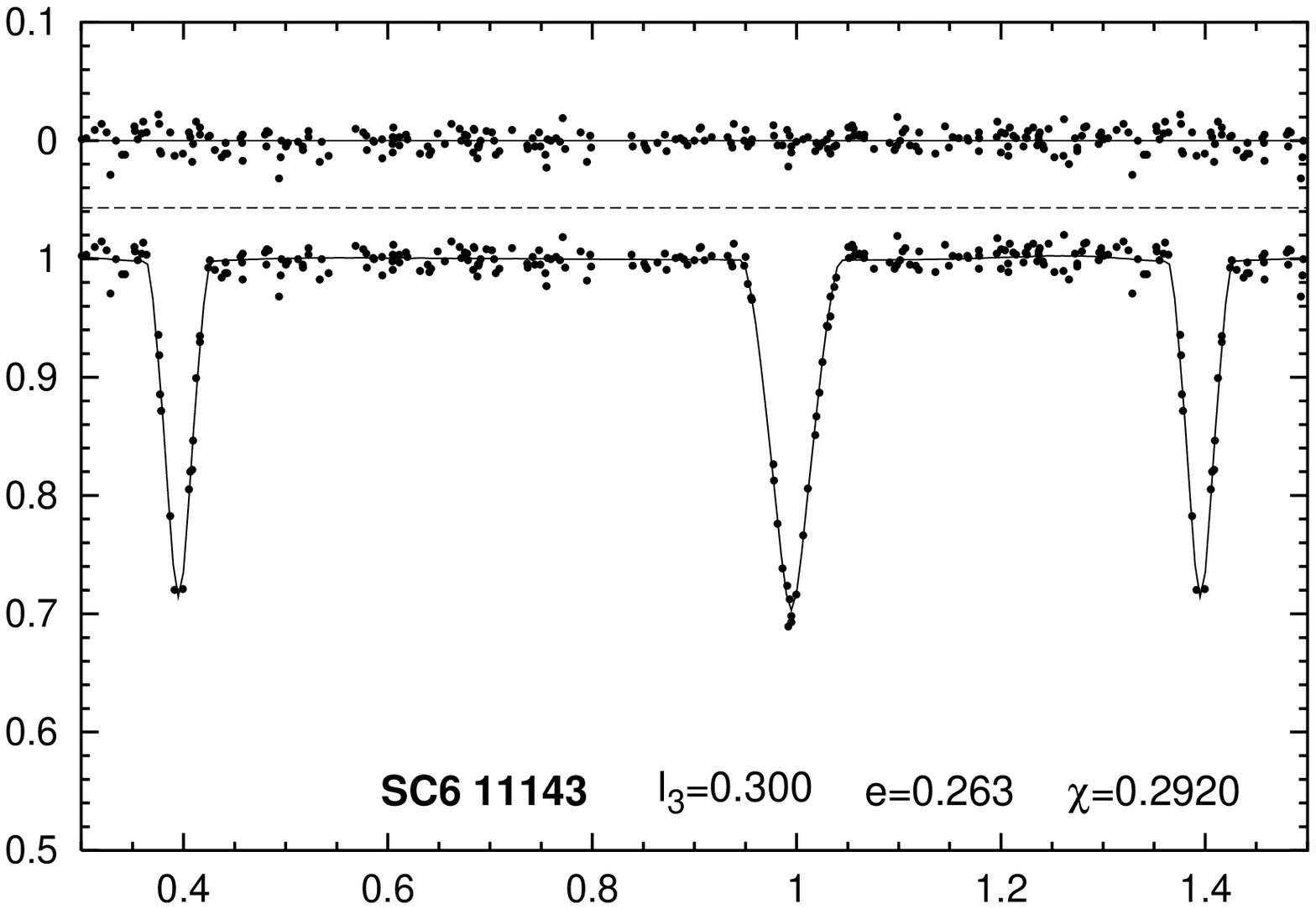}}
\mbox{}\\[-0.55cm]
  \resizebox{\linewidth}{!}{\includegraphics{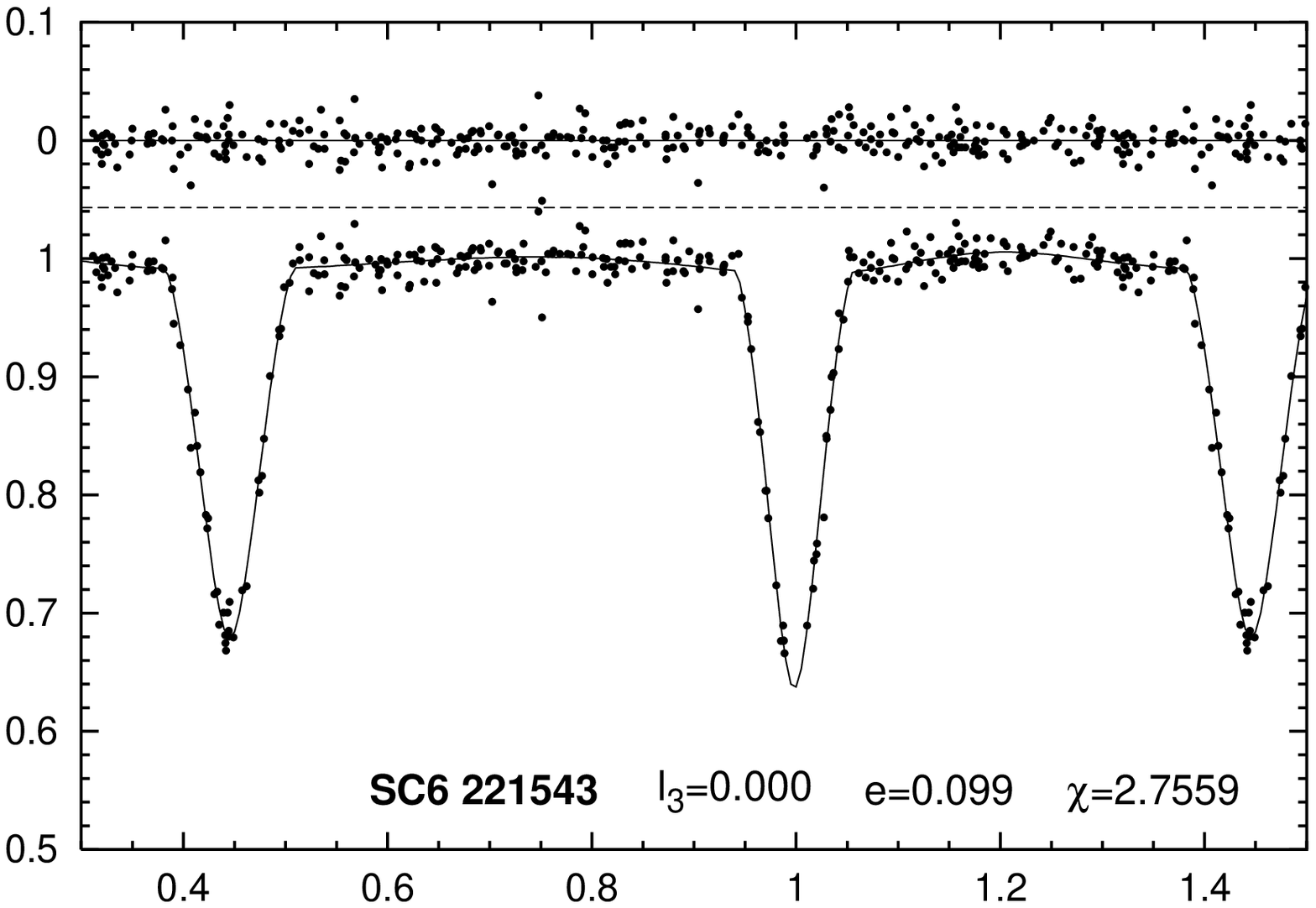}}
\mbox{}\\[-0.55cm]
  \resizebox{\linewidth}{!}{\includegraphics{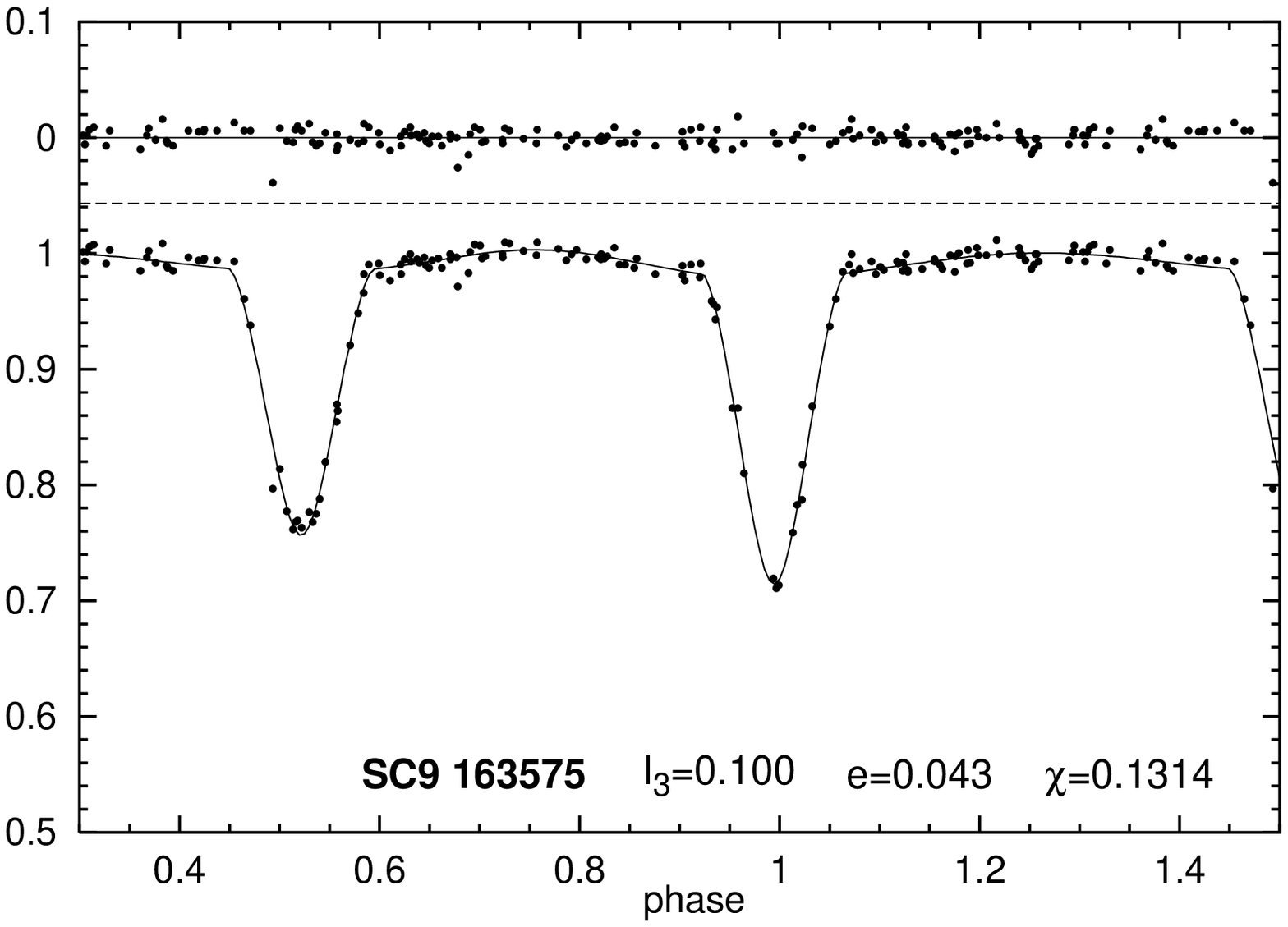}}
\end{minipage}
\caption{The light curves solutions and their residuals for the systems 
from the A subset.}
\label{fig:fitA}
\end{figure*}

%------------------------------------------------------------
For each system from A group the automated fitting scheme described in \S 3.1 was 
employed. Only in three cases: SC6 67221, SC6 221543 and SC6 272453 was the convergence achieved
by adjusting simultaneously all free parameters. The Figure~\ref{fig:fitA} shows the best-fitting model light curves, 
together with $O-C$ residuals, whose photometric parameters 
are given in Table~\ref{tab:solA}. All arguments of the periastron refer
to the epoch 1999.0.

For all systems from the A subset reasonable solutions were found assuming $l_3 = 0$. 
However, in the case of SC6 11143 and SC9 163575 we expect that some amount
of the third light is present in the system because of the blending
with another eclipsing variable. In order to estimate third light component 
an additional grid of solutions for different values of $l_3$ was calculated in the range
of 0.05-0.40 with a step of 0.05. For SC6 11143 the best solution was found at $l_3=0.30$, while for SC9 163575
 -- at $l_3=0.10$. It is worth noting that the solutions found with $l_3=0$ for these
systems have almost no systematic trends in O-C's residuals, thus the intrinsic variability
of the blend is the only sign of the third light component. 

The solutions for SC6 67221 converged very rapidly and had the smallest 
residuals. Only slight systematic residuals are visible near phase 0.04 
(just after the egress from primary minimum), most probably due to the strong mutual reflection
effect. Similar but even stronger systematic effect is visible near first quadrature
($\phi\approx 0.11$) in SC5 129441. Because of the presence
of the proximity effects  both systems were additionally investigated in order
to estimate possibly the third light contribution by adjusting $l_3$ as a free
parameter.
In each case synthetic light curves produce a tiny improvement only for 
unphysical values $l_3$$<$0, but such solutions were invariably
unstable. The conclusion is that these binaries show virtually insignificant
third light contribution what is consistent with $l_3=0$. For a brief 
discussion of both systems see~\S~5.1.2.

Small residuals and relatively quick convergence characterize 
also SC3 139376 and SC9 163575, whereas, the largest residuals are visible 
in SC5 221543 mostly due to a big intrinsic scatter in observations.     

Having calculated $e$ and $\omega$
we could estimate the rate of apsidal motion for two binaries from the sample. The change of displacement of
minima $\Delta\Phi$ of an eccentric binary with central eclipses ($i$=90\degr) is given by
(see e.g.~Kopal 1959, Chapter 6, equations (9-8), (9-9)): 
\begin{equation}
\Delta\Phi = \frac{e \sin{\omega}\,\dot{\omega}}{\sqrt{1-e^2}}. \label{apsid}
\end{equation}
However for our purpose we can still use this formula also for inclination
close to 90\degr provided the timescale of a displacement change
is small with respect to an apsidal period $U$.
The change of displacement per one orbital revolution ($P$) is simply $\alpha$
(Eq.~\ref{shift}) so the rate of advance of periastron is
approximated by:
\begin{equation}
\dot\omega \approx \frac{365\, \alpha \sqrt{1-e^2}}{P\, e \sin{\omega}} \,\,\,
[\,\degr\!/{\mathrm yr}]. \label{rate}
\end{equation}
Substituting numerical values we get 5\fdg
8/yr and 1\fdg 8/yr for SC3 139376 and SC5 311566, respectively.

\begin{figure*}
\begin{minipage}[ht]{0.5\linewidth}
\hspace*{-0.14cm}\resizebox{1.03\linewidth}{!}{\includegraphics{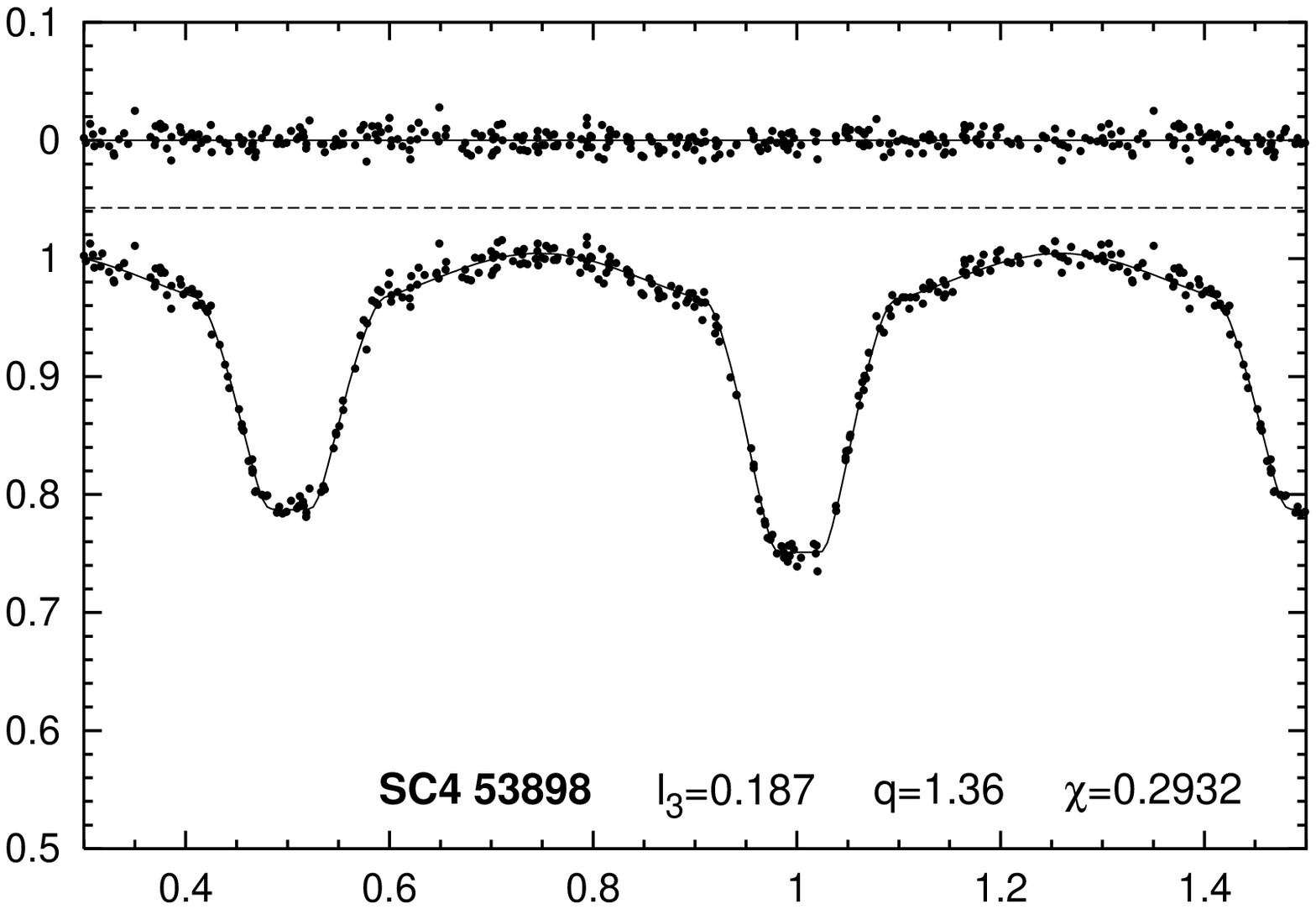}}
\mbox{}\\[-0.55cm]
  \resizebox{\linewidth}{!}{\includegraphics{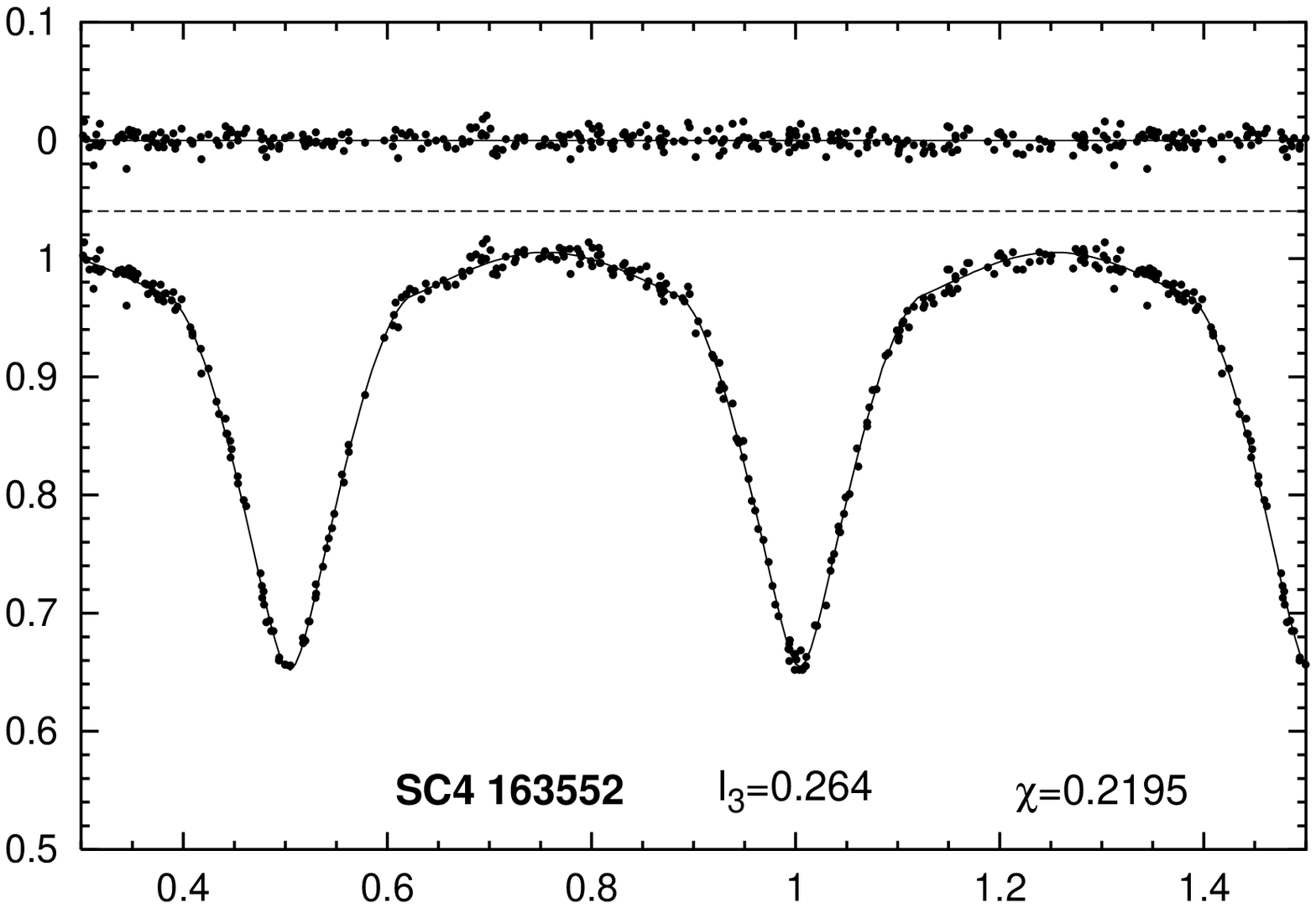}}
\mbox{}\\[-0.55cm]
  \resizebox{\linewidth}{!}{\includegraphics{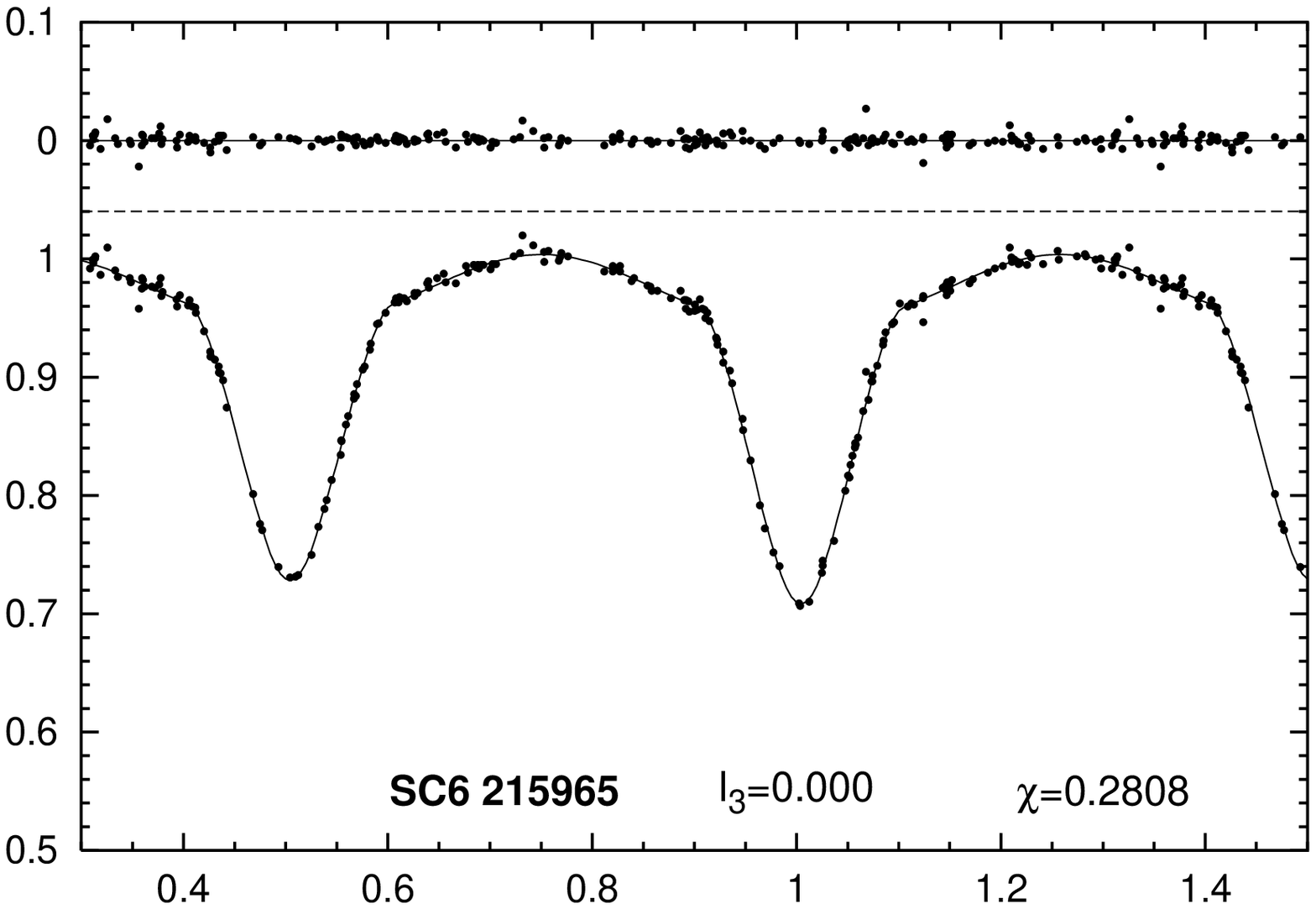}}
\mbox{}\\[-0.55cm]
  \resizebox{\linewidth}{!}{\includegraphics{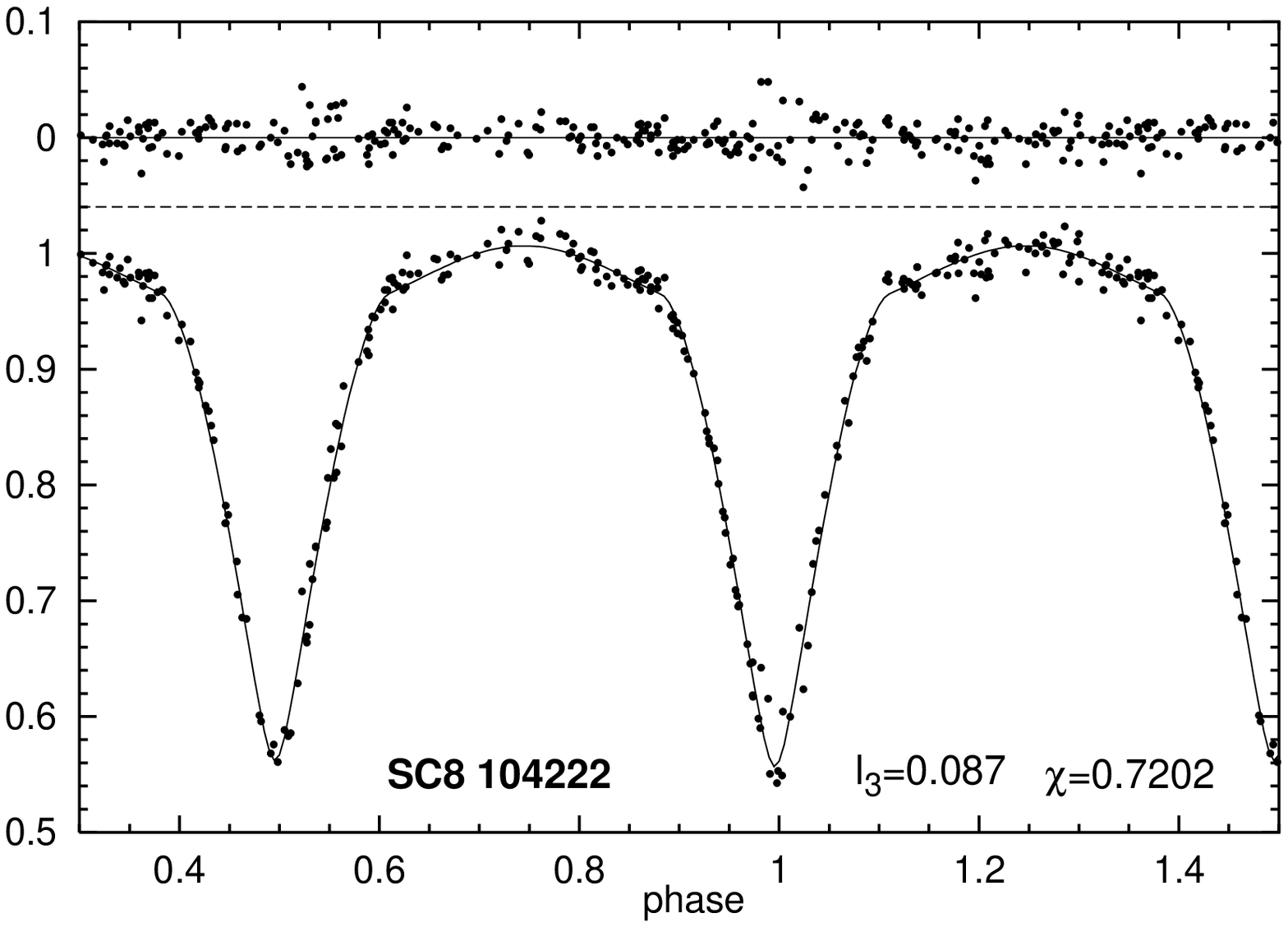}}
\end{minipage}\hfill
\begin{minipage}[h]{0.5\linewidth}
  \resizebox{\linewidth}{!}{\includegraphics{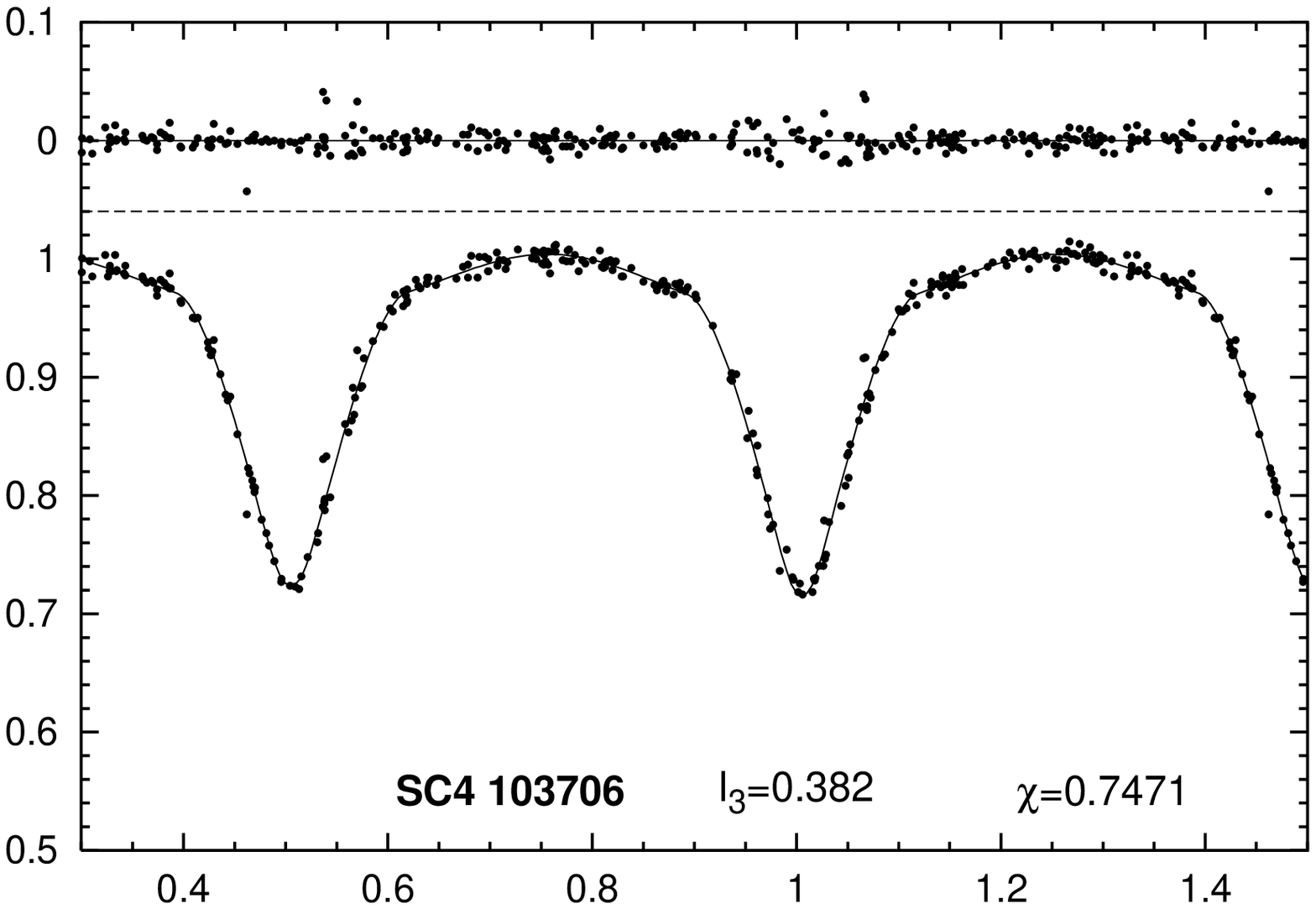}}
\mbox{}\\[-0.55cm]
  \resizebox{\linewidth}{!}{\includegraphics{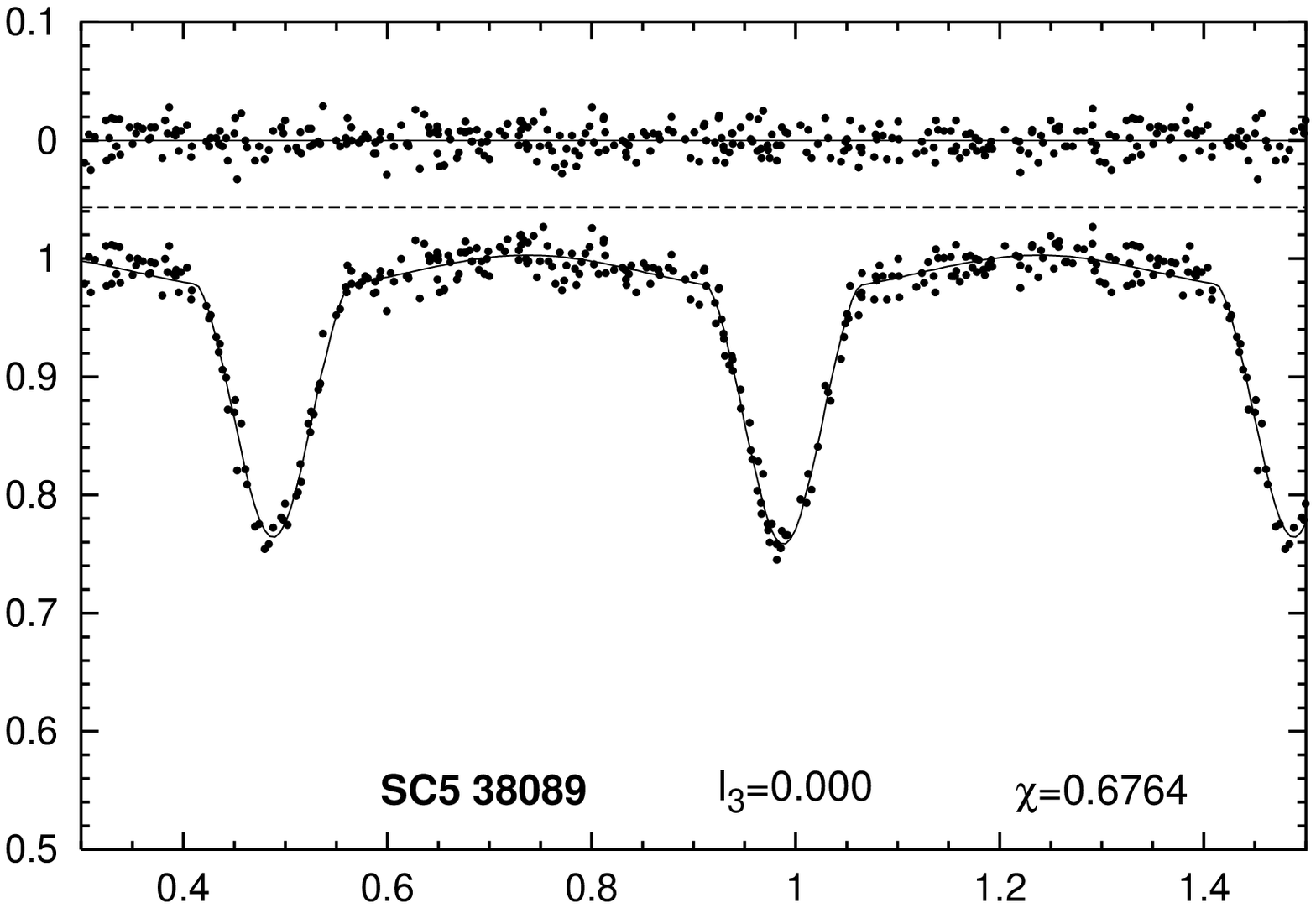}}
\mbox{}\\[-0.55cm]
  \resizebox{\linewidth}{!}{\includegraphics{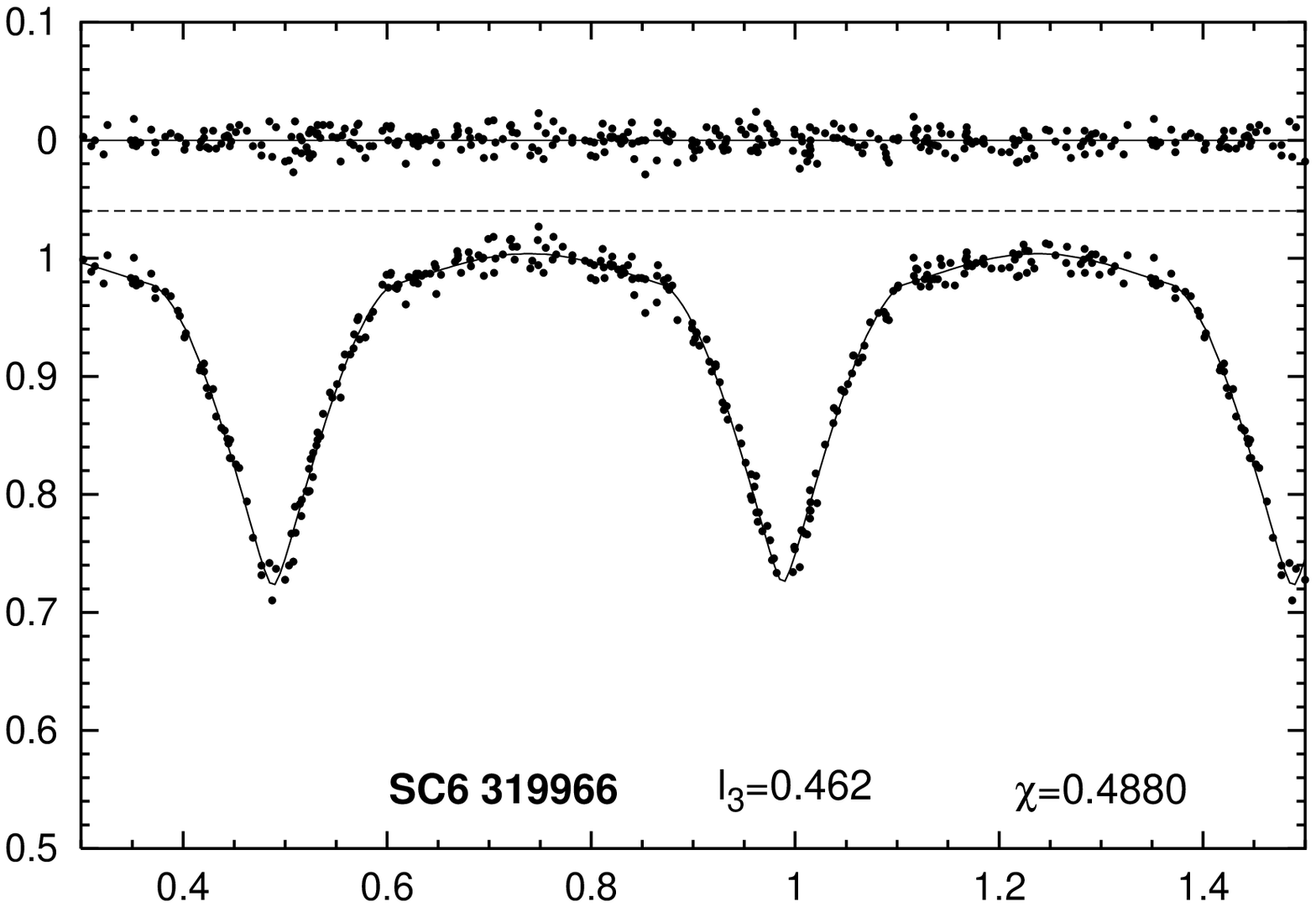}}
\mbox{}\\[-0.55cm]
\hspace*{-0.14cm}\resizebox{1.03\linewidth}{!}{\includegraphics{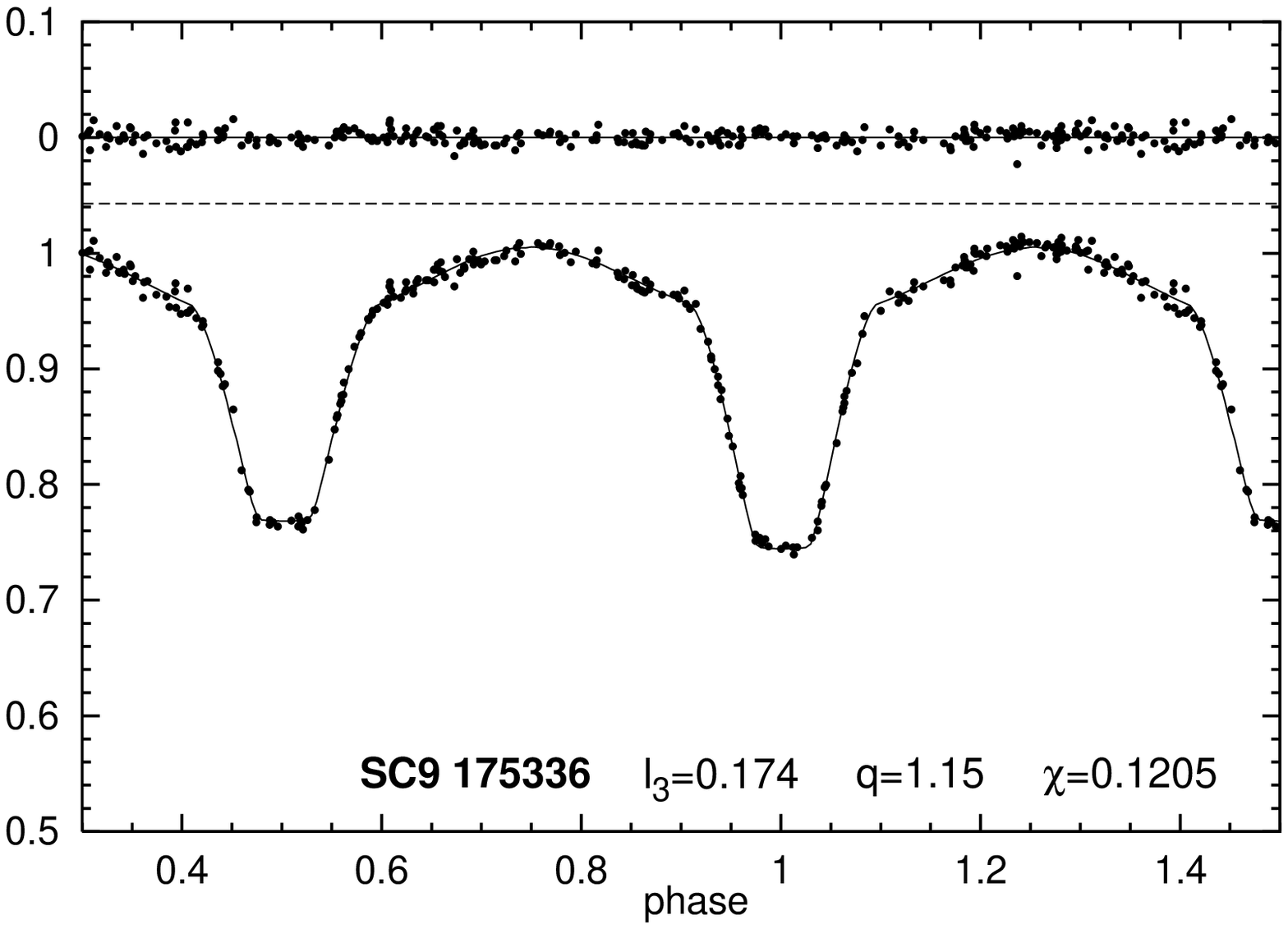}}
\end{minipage}
\caption{The light curves solutions and their residuals for the systems 
from the B subset.}
\label{fig:fitB}
\end{figure*}

\begin{figure*}
\begin{minipage}[ht]{0.5\linewidth}
  \resizebox{\linewidth}{!}{\includegraphics{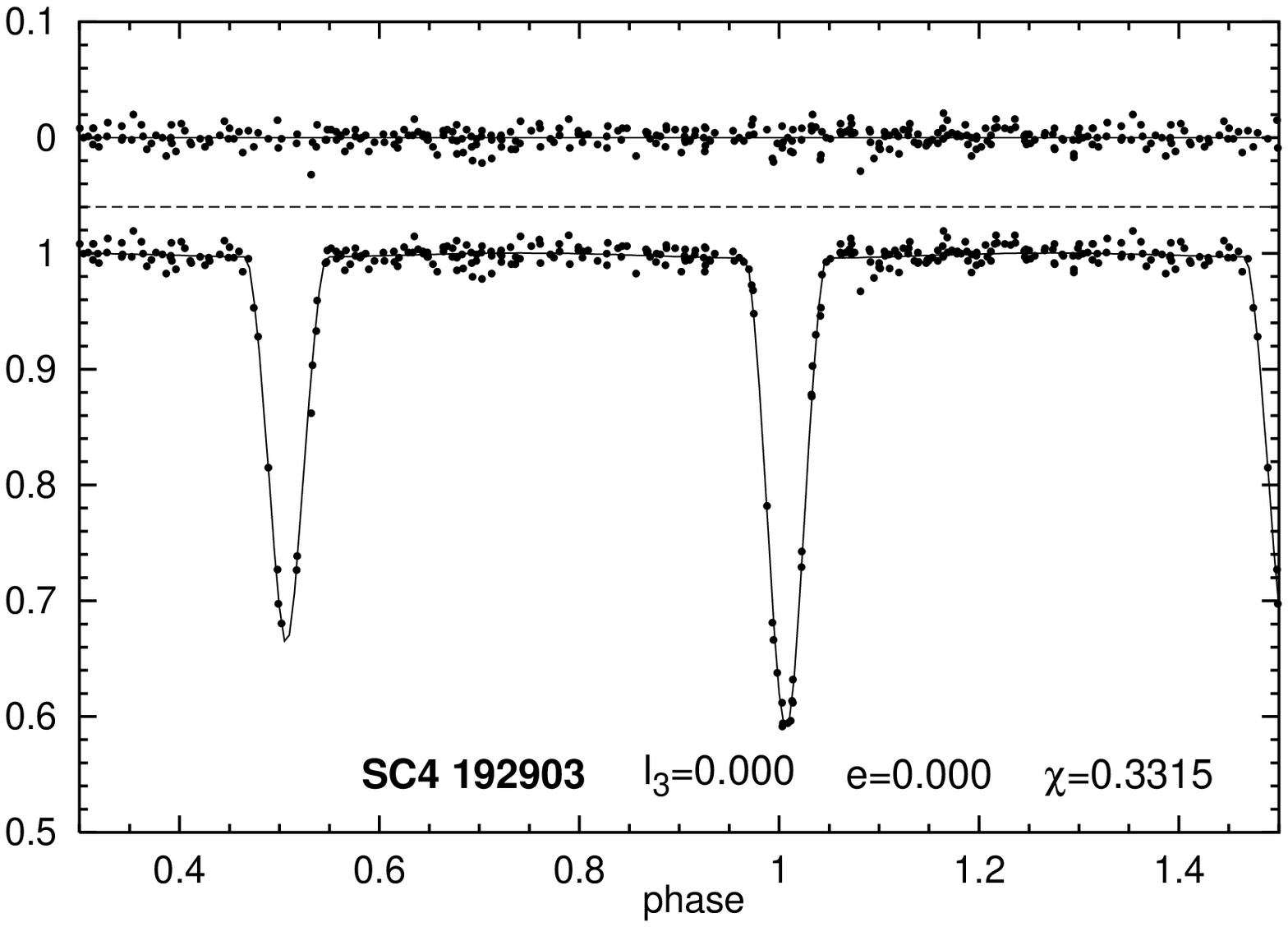}}
\end{minipage}\hfill
\begin{minipage}[h]{0.5\linewidth}
  \resizebox{\linewidth}{!}{\includegraphics{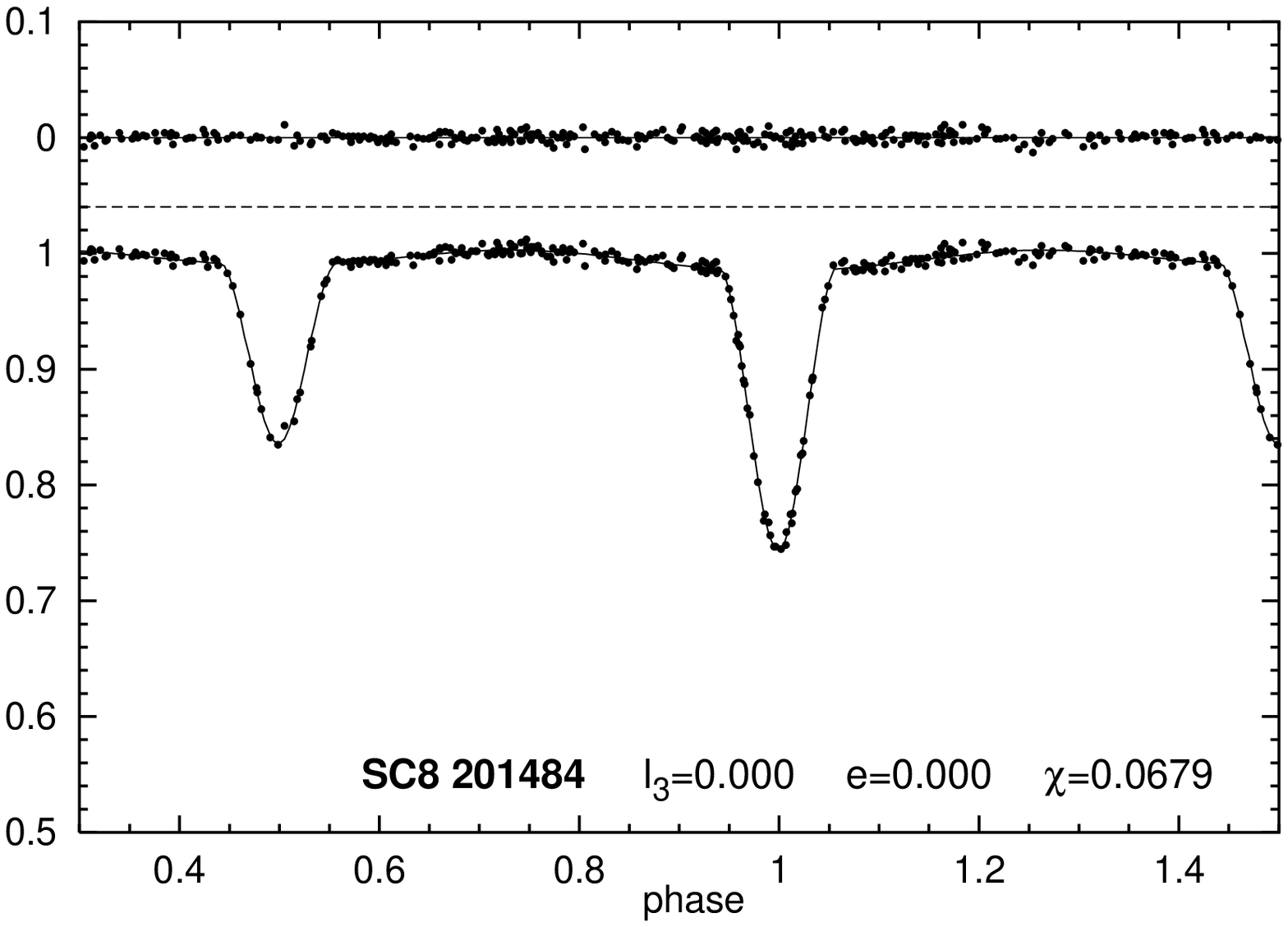}}
\end{minipage}
\caption{The light curves solutions and their residuals for the systems 
from the C subset.}
\label{fig:fitC}
\end{figure*}

\subsection{Solutions for B and C subsets}
%-------------------------------------------------
The photometric parameters for the stars from B and C subsets are presented in
Table~\ref{tab:solA}. Appropriate 
light curve solutions are
shown in Figs.~\ref{fig:fitB} and~\ref{fig:fitC}. 
In contrast to the A subset,
almost all solutions from B subset (beside SC6 215956) were found
only by adjusting the third light $l_3$ in the system. 
There is a simple explanation of this effect - see a discussion
in section~\ref{l3}. SC6 215956 was also investigated for
possible third light contribution but values
of $l_3$ suggested by WD program were invariably negative.   
SC6 215965 has very small residuals and the solution for this system
converged rapidly. Also it seemed to be most stable of all in B subset.

The solution for SC5 38089 was obtained using the spectroscopic mass ratio $q=1.12$ and 
more appropriate temperatures ($T_{1,2}\approx 29000$ K) 
determined for this binary by Harries et al.~(2003). Third light contribution in this system 
is spurious although some improvements to the fits can be obtained for $l_3>0.2$.  

In spite of the small residuals in SC9 175336 its solution was extremely unstable
and exhibited large systematic $O-C$ deviations, especially near
external contacts. The tests showed that the
reason was the sensitivity of the light curve to the mass ratio $q$
as the larger and cooler component seemed to
be noticeably more evolved than the hotter one. The same problems
arose for SC4 53898 which seems to be similar in many respects
to the former, although its light curve is less sensitive for $q$. 
Both systems show complete eclipses which allow for  
more accurate determination of photometric parameters 
including the photometric mass ratio.  
Thus I allowed for adjusting of $q$ for both systems in detached
configuration. The convergence was achieved and a stable solution obtained: 
in the case of SC9 175336 for $q=1.15\pm 0.05$ and in the case of SC4 53898 for 
$q=1.36\pm 0.12$.   

The solutions for two systems from C subset converged very rapidly.
No third light $l_3$ was adjusted in both cases because of the lack of any
systematic residuals in the solutions (Fig.~\ref{fig:fitC}). Moreover 
the solution for SC8 201484 has the smallest residuals from all of the subsets.  

\section{Physical parameters}
%-------------------------------------
\label{physics}
We cannot directly calculate absolute dimensions and masses without knowing spectroscopic orbit.
However it is possible to estimate stellar parameters using the inversion of the method of
parallaxes of eclipsing binaries. The method of parallaxes was elaborated
by Dworak (1974) who based it on Gaposchkin (1938) algorithm. We assume
that we {\it know} distance to SMC and we use it to calculate absolute
dimensions by scaling the system to obtain the observed flux at Earth. 
The calculations are quite straightforward. For early B or O type stars the
Rayleigh-Jeans approximation may be used in visible and near-infrared region 
of the spectrum so total $V$ or $I$ luminosity 
of each star is a linear function of temperature \( L_{1,2} \sim T_{1,2} \). 
Using this approximation we can 
disentangle the temperatures $T_1$ and $T_2$ of both components: 
\begin{equation}
T_1=T_0 \frac{1+L^I_{21}}{1+L^I_{21}(T_2/T_1)},  
\end{equation}
where \(L^I_{21} = L^I_{2}/L^I_{1} = l_2/l_1\) is the luminosity ratio
of the components known from the light curve solution (Table~\ref{tab:solA}) 
and $T_0$ is the "mean" temperature corresponding to the
total $(B-V)_0$ colour of the system. The temperature $T_2$
is then directly calculated from the temperature ratio. Initially $(B-V)_0$ was found for each
binary assuming total, foreground and internal, mean reddening to SMC
$E(B-V)=0.087$ (Massey et al.~1995). 
The foreground reddening of $E(B-V)=0.037$ was assumed
(Schlegel, Finkebeiner \& Davis~1998) with the ratio of 
the selective to the total extinction $R=3.1$, while for the internal
reddening I assumed $R=2.7$ (Bouchet et al.~1985).   
The resulting temperatures $T_{1,2}$ were used to calculate approximate intrinsic
colours $(B-V)_0$ and bolometric corrections $BC_{1,2}$ of both components via Flower's (1996) 
calibration for main sequence stars.
The "standard" mean value \(m-M=18.9\) of the distance modulus ($DM$) for SMC was assumed. 
The semi-major axis $A$ can be then computed by the following formula:
\begin{eqnarray}
\;\;\log{A} &=& 0.2\, DM - \log{r_1} - 2 \log{T_1} + 7.524 + \nonumber \\ 
        & & 0.2\, (M^V_\odot - BC_1 - V_1 + R\, E(B-V)), \label{axis}
\end{eqnarray}
where $M^{V}_{\odot}$ is the $V$ absolute magnitude of the Sun (assumed +4.75), 
$V_1$ is the out-of-eclipse $V$ magnitude of the primary (calculated 
assuming $L^V_2/L^V_1 \approx l_2/l_1$ and that $l_3$ contribution in $V$ band 
is comparable to that in $I$ band). The total mass of the system results
simply from application of Kepler's third law. The absolute scale
of the system gives immediately the bolometric luminosity $L$ of each component.

Now we proceed to the second step of the method.
We can represent the luminosity of the star as a function of its mass,
which is well-known as Eddington's relation. This relation works only for main
sequence stars, but it should be kept in mind that for an early spectral
type star (massive and luminous at the same time) its luminosity 
is not a strong function of the age. The dependence is rather weak and
such star tends to evolve from ZAMS along almost horizontal line on HR diagram.
Moreover, no binary from the sample (beside two red systems from C subset) seems to
contain considerably evolved components. 

\begin{figure}
\begin{minipage}{\linewidth}
  \resizebox{1.02\linewidth}{!}{\includegraphics{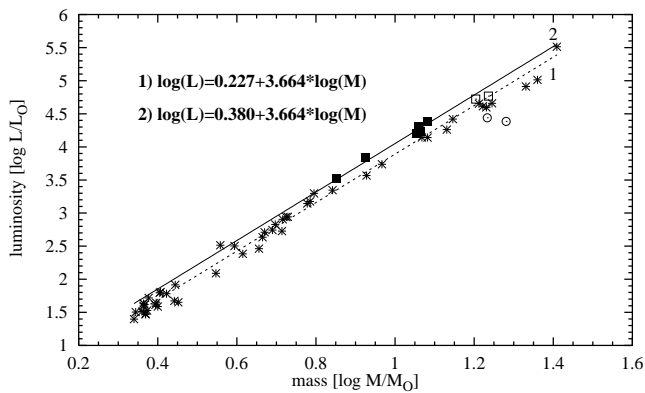}}
%  \resizebox{0.5\linewidth}{!}{\includegraphics{mas_jasn.ps}}
\end{minipage}
\caption{Mass-luminosity relation for Galactic main
sequence stars (asterisks) given by equation 1 and LMC stars (filled squares)
-- equation 2. Both components of two SMC detached systems: SC6 215965 (open squares) and SC5 38089 (open circles) from paper
Harries et al.~(2003) are also marked. Note that the components of SC6 215965 are placed almost on the predicted M-L relation, 
while the components of SC5 38089 do not seem to concur with this mass-luminosity relationship.}
\label{fig:law}
\end{figure}

Fig.~\ref{fig:law} presents the mass-luminosity relation. The data for
massive ($M>2.2 M_\odot$) Galactic 
stars were taken from compilation of detached, double lined eclipsing binaries done by 
Pols et al.~(1997 and references therein) and for three LMC eclipsing systems from Ribas et
al.~(2000), Fitzpatrick et al.~(2001) and
Ribas et al.~(2002). The inclusion of LMC stars gives an opportunity to 
check how this relation depends on the metallicity. Indeed the Galactic
stars seems to have a little smaller luminosity
then LMC stars (having [Fe/H]$\approx -0.35$)
at a given mass. For a Galactic star the mass-luminosity $M-L$ relation may be
approximated by a simple linear function in wide range of
masses ($2.2 M_\odot <M<20 M_\odot$):
\begin{equation}
\log{L} = 3.664(47)\; \log{M} + 0.227(38), \label{mas:gal}
\end{equation}
i.e. \(L \sim M^{3.66}\).
The sample of LMC stars do not cover such wide range of masses and I assumed
that the slope of the $M-L$ relation is the same like for the Galactic stars. 
Then the data can be fitted by:
\begin{equation}
\log{L} = 3.664(47)\; \log{M} + 0.380(27)\,. \label{mas:lmc}
\end{equation}
Above equation (labeled 2 in Fig.~\ref{fig:law}) gives the
approximate $M-L$ relation for LMC stars. This relation can be compared with 
a mass determination for two detached eclipsing systems in SMC done by Harries et al.~(2003) 
-- see Fig.~\ref{fig:law}. Components of SC6 215965 lie very close to the predicted 
M-L relation, but there are some problems with components of SC5 38089: especcially 
a more massive component of 38089 seems to be a less luminous one. As the physical 
parameters of 215965 are more precisely determined than in the case of 38089  
I decided to adopt the mass-luminosity relation
given by Eq.~\ref{mas:lmc} for SMC stars. Relying on this assumption we can check
the consistency between parameters found from the flux scaling
(Eq.~\ref{axis}) and from $M-L$ relation (Eq.~\ref{mas:lmc}).

\begin{table*}
\begin{minipage}{\linewidth}
\caption{Physical parameters}\label{tab:par}
\tabcolsep2pt
\begin{tabular}{@{}|l|c|c|c|c|c|c|c|c|c|c|c|c|c|@{}}
\hline
\hline
    &               &\multicolumn{2}{c}{Mass}          &Mass
&\multicolumn{2}{c}{Mean radius}       &     &  &\multicolumn{2}{c}{$\;\;$Temperature$\;\;$}&\multicolumn{2}{c}{Luminosity}& \\
Object&Spectrum&$\frac{M_1}{M_\odot}$&$\frac{M_2}{M_\odot}$&ratio
$q$&$\frac{R_1}{R_\odot}$&$\frac{R_2}{R_\odot}$&$\log{g_1}$&$\log{g_2}$
&\hspace{0.2cm}$T_2$ & $T_1$         & $\log{L_1}$& $\log{L_2}$&$E(B$--$V)$ \\\hline
\multicolumn{14}{c}{The A subset}\\
SC3  139376$^\ast$&B0V+B0V 		                         	& 15.4$\pm$2.0 & 16.3$\pm$2.1&  1.06$\pm$0.11&  7.8$\pm$0.9&  8.6$\pm$1.0&  3.84&  3.79& $\;\,$31500$\pm$1100& $\;\,$31700$\pm$1100& 4.731& 4.823 &0.128   \\
SC5  129441$^\ast$$^\dagger$ & B3III+B3III                &  $\;\,$7.1$\pm$0.6 &  $\;\,$7.2$\pm$0.6&  1.01$\pm$0.07&  6.5$\pm$0.4&  6.7$\pm$0.4&  3.67&  3.65& 16900$\pm$400& 16900$\pm$400& 3.486& 3.506 &0.094   \\
SC5  311566 & O9V+O9V                                	  & 11.9$\pm$1.6& 11.0$\pm$1.5&  0.92$\pm$0.10&  4.2$\pm$0.5&  3.9$\pm$0.5&  4.28&  4.30& $\;\,$34200$\pm$1300& $\;\,$32900$\pm$1200& 4.324& 4.199 &0.180   \\
SC6   11143$^\dagger$&B2V+B2V                             &  $\;\,$7.5$\pm$0.8&  $\;\,$6.6$\pm$0.7&  0.88$\pm$0.08&  4.3$\pm$0.5&3.6$\pm$0.4& 4.04&  4.14& 21600$\pm$600& 21000$\pm$600& 3.565& 3.367 &0.103   \\
SC6   67221$^\ast$ &B2IV+B1IV 							            & 10.0$\pm$1.1&11.1$\pm$1.2&  1.11$\pm$0.09&  7.5$\pm$0.6&  6.8$\pm$0.6&  3.69&  3.82& 21700$\pm$600& 25100$\pm$700& 4.052& 4.218 &0.153   \\
SC6  221543 & B2V+B2V			                               &  $\;\,$8.2$\pm$1.0&  $\;\,$7.4$\pm$1.0&  0.90$\pm$0.09&  4.8$\pm$0.5&  4.3$\pm$0.5&  3.99&  4.03& 22400$\pm$600& 21500$\pm$600& 3.715& 3.554 &0.027   \\
SC6  272453 &B2V+B2IV								                   &  $\;\,$7.9$\pm$0.9&  $\;\,$8.4$\pm$1.0&  1.06$\pm$0.09&  5.1$\pm$0.6&  6.1$\pm$0.6&  3.93&  3.79& 21400$\pm$600& 20500$\pm$500& 3.680& 3.775 &0.065   \\
SC9  163575 & O5+O7                                  	  & 16.9$\pm$2.4& 14.6$\pm$2.0&  0.86$\pm$0.10&  4.9$\pm$0.6&  4.7$\pm$0.6&  4.29&  4.26& $\;\,$43300$\pm$1800& $\;\,$38600$\pm$1500& 4.875& 4.640 &0.073   \\
SC10  37223 & B1V+B1V                                	  &  $\;\,$8.4$\pm$1.0&  $\;\,$7.8$\pm$0.9&  0.93$\pm$0.08&  4.0$\pm$0.4&  3.9$\pm$0.4&  4.15&  4.15& 25100$\pm$700& 23900$\pm$650& 3.766& 3.648 &0.064   \\
\multicolumn{14}{c}{The B subset}\\
SC4   53898$^\ast$ &B1V+B2IV                       			 &  $\;\,$6.5$\pm$0.6&  $\;\,$8.9$\pm$0.9& $\;\,$1.36$\pm$0.12$^\ddag$&  2.9$\pm$0.2&  5.4$\pm$0.4&  4.34&  3.93& 24400$\pm$700& 22200$\pm$600& 3.415& 3.801 &0.122   \\
SC4  103706$^\dagger$ & B0V+B0V                           & 10.7$\pm$1.3&  $\;\,$9.1$\pm$1.1&  0.86$\pm$0.08&  4.9$\pm$0.4&  4.0$\pm$0.4&  4.09&  4.21& $\;\,$28700$\pm$1000& 27700$\pm$800& 4.162& 3.917 &0.085   \\
SC4  163552$^\dagger$ & B1V+B1V                           &  $\;\,$9.3$\pm$1.0&  $\;\,$9.2$\pm$1.0&  0.98$\pm$0.09&  5.0$\pm$0.5&  4.9$\pm$0.5&  4.01&  4.03& 24700$\pm$700& 24800$\pm$700& 3.922& 3.897 &0.216   \\
SC5   38089 & B0V+B0V                                     & 11.7$\pm$2.0& 11.8$\pm$1.9&  1.01$\pm$0.12&  5.3$\pm$0.9&  5.5$\pm$0.8&  4.10&  4.07& $\;\,$29700$\pm$1000& $\;\,$29400$\pm$1000& 4.291& 4.302 &0.047   \\
SC6 215965$^\ast$$^\dagger$&B0V+B1IV 									& 14.6$\pm$1.1& 14.8$\pm$1.1&  1.02$\pm$0.05&  9.0$\pm$0.7&$\!\!$10.9$\pm$0.8&  3.70&  3.54& 28100$\pm$900& 25900$\pm$800& 4.653& 4.677 &0.039   \\
SC6  319966$^\dagger$ & B2V+B2V                           &  $\;\,$6.2$\pm$0.6&  $\;\,$6.1$\pm$0.6&  0.98$\pm$0.07&  3.4$\pm$0.3&  3.3$\pm$0.3&  4.16&  4.19& 20700$\pm$500& 20800$\pm$500& 3.286& 3.259 &0.085   \\
SC8  104222 & B1V+B1V                                	  &  $\;\,$8.4$\pm$1.0&  $\;\,$8.7$\pm$1.1&  1.03$\pm$0.10&  4.6$\pm$0.6&  4.9$\pm$0.5&  4.05&  4.01& 23900$\pm$650& 23800$\pm$650& 3.784& 3.826 &0.113   \\
SC9  175336$^\ast$ &B1V+B1IV                  		      & 10.0$\pm$1.0& 11.4$\pm$1.1& $\;\,$1.15$\pm$0.05$^\ddag$&  4.7$\pm$0.4&  8.7$\pm$0.6&  4.09&  3.62& 25200$\pm$700& 24300$\pm$700& 3.901& 4.371 &0.039   \\
\multicolumn{14}{c}{The C subset}\\
SC4  192903 &F9IV+G2IV                				     		   &  $\;\,$1.5$\pm$0.3&  $\;\,$1.6$\pm$0.3&  1.1$\pm$0.2&  15$\pm$2&  19$\pm$2&  2.26&  2.06&  $\;\,$5900$\pm$100&  $\;\,$5400$\pm$100& 2.372& 2.472 &0.087   \\
SC8  201484 &  G0III+G7III                                 &  $\;\,$1.9$\pm$0.3&  $\;\,$1.9$\pm$0.3&  1.0$\pm$0.2&  36$\pm$2&  48$\pm$3&  1.59&  1.36&  $\;\,$5700$\pm$100&  $\;\,$5000$\pm$100& 3.088& 3.116 &0.087   \\
\multicolumn{14}{c}{LMC stars}\\
EROS 1044   & B2V+B2IV								  	             &  $\;\,$7.1$\pm$0.8&  $\;\,$8.7$\pm$0.9&  1.22$\pm$0.10&  4.2$\pm$0.4&  6.7$\pm$0.6&  4.04&  3.73& 21400$\pm$600& 20500$\pm$500& 3.524& 3.844 &0.065   \\      
HV    982   & B1V+B1V          		       		              & 10.4$\pm$1.1& 10.6$\pm$1.1&  1.02$\pm$0.08&  7.0$\pm$0.5&  7.6$\pm$0.5&  3.77&  3.70& 23300$\pm$650& 22800$\pm$650& 4.111& 4.143 &0.081   \\      
HV   2274   & B2IV+B2IV           			                & 12.5$\pm$1.3& 12.0$\pm$1.2&  0.96$\pm$0.07&$\!\!$10.0$\pm$0.7&  9.2$\pm$0.7&  3.54&  3.60& 23100$\pm$650& 23200$\pm$650& 4.407& 4.341 &0.121   \\\hline
\end{tabular}
\end{minipage}
\medskip
\begin{minipage}{\linewidth}
{\footnotesize Note: all parameters were estimated assuming
$m-M=18.9$ for SMC stars and $m-M=18.5$ for LMC stars. The spectrum was estimated
from $T$ and $\log{g}$ and, in the case of LMC
stars, adopted from the literature. The stars chosen in the present paper as likely distance
indicators are marked by an asterisk. The stars indicated by Wyithe \&
Wilson (2001, 2002) are marked by a dagger. A double
dagger indicates a photometric mass ratio derived from the light curve analysis.}
\end{minipage}
\end{table*}

Such treatment allows for independent determination of the
unknown reddening in the direction to the particular binary and thus
a reasonable $(B-V)_0$ intrinsic colour can be obtained. 
Although the mean foreground reddening toward SMC 
is quite low some large random internal reddenings (up to 0.2) 
were reported for individual B type stars (Larsen, Clausen \& Storm 2000).
It is obvious (e.g.~Flower's 1996 calibration) that for early type stars       
the temperature is a very strong function of the $(B-V)_0$ colour and a small
change of this colour causes a large change of the temperature. 
As the temperature enters Eq.~\ref{axis} at a relative high power even a small 
error in determination of reddening $E(B-V)$ produces large errors
in luminosities and masses.
Let us return to the problem of the determination of physical parameters.
%Assuming that both stars should follow Eq.~{\ref{mas:lmc} 
We can compute the expected mass ratio $q$ as follows: 
\begin{equation}
\log{q}=(\log{L_2}-\log{L_1})/3.664, \label{ratio}
\end{equation}
where $L_{1,2}$ are the absolute luminosities computed by using Eq.~\ref{axis}.
Then the masses of individual components 
$M_{1,2}$ are calculated from Eq.~\ref{mas:lmc}. If there is a consistency 
between the flux scaling and $M-L$ relation a quantity $\beta$:
\begin{equation}
\beta = 0.5\, \left(\log{L_1}+\log{L_2}-3.664\,(\log{M_1}+\log{M_2})\right) \label{alfa}   
\end{equation}
should be equal or close to the free term in Eq.~\ref{mas:lmc}
i.e.~0.38. I~have allowed $\beta$ parameter to vary in the range of
0.36-0.40. If $\beta$ were in allowed range the consistency was
achieved, the $E(B-V)=0.087$ was approved as the final reddening and calculations 
stopped. If not, the iterative determination of the reddening were done. 
For $\beta$ greater than 0.40 the correction of $-0.001$ were added to
$E(B-V)$, while for $\beta$ smaller than 0.36 the correction of $+0.001$ was 
applied and all calculations were repeated for this new value of
reddening, until the $\beta$ fell in the allowed range.

The above method was tested
on three LMC eclipsing binaries which have their physical
parameters known with high accuracy. As a free parameter I assumed
distance modulus to the LMC. The best "fit" to the original data were
obtained for $m-M=18.5$ what is a value still accepted as close to the
true one (e.g.~Gibson 1999) and marginally consistent with $DM$ obtained
using eclipsing binaries (Ribas et al.~2002). The comparison between original data and
parameters given in Table~\ref{tab:par} (lowest panel) show that the "fit" is excellent --
the method allows for consistent and accurate estimation of masses, 
sizes, temperatures, luminosities and reddenings of all three systems.
It may serve as a proof that such an approach of estimation of
physical parameters works well and may give fast and 
proper information about the absolute dimensions for large samples
of eclipsing binaries -- see \S~\ref{fast} for a discussion. It is worth to note
that the initial value of reddening is irrelevant to the method and the
same results we obtain adopting another initial $E(B-V)$.

Table~\ref{tab:par} gives the final estimation of the physical parameters 
of each component of binaries in the sample. The spectral
type and luminosity class of each star was
estimated from $T$ and $\log{g}$. 
SC4 53898 and SC9 175336 were treated in a slightly different way. The mass
ratio for both systems was not calculated from Eq.~\ref{ratio} but 
was assumed to be equal to the photometric mass ratio determined
from the light curve solution (see \S~3.4). The further calculations were
done in a similar way like for the rest of the sample.  
For two binaries from C subset
which host evolved and cool components the reddening was assumed to be equal
to the initial value and just the simple flux scaling was applied.

The temperatures $T_1$ from Table~\ref{tab:par} in a few cases
(SC4 163552, SC5 311566, SC9 175336) significantly differ from initial 
$T_1$ for which light curve solutions were found. 
Additional tests show, however, that photometric parameters recalculated
for the temperatures resulted from the flux scaling were consistent
with those from Table~\ref{tab:solA} within quoted errors. 
 
\section{Distance indicators}
%-------------------------------------
Below the discussion of individual eclipsing binaries as distance
indicators to SMC is presented. Six candidates were chosen
according to the following criteria:
the presence of complete eclipses, small residuals, small and well
determined third light contribution and the spectral type.
%Their positions relative to the Small
%Magellanic Cloud were indicated in Fig.~\ref{fig:cand}. 
Two binaries from this subset were also advocated to be possible distance 
indicators by WW1. Another four systems which were selected by WW1 and WW2
were also briefly discussed in a separate subsection, where reasons against their 
selection were given and additionaly SC5 38089 used recently for distance determination 
to SMC is also discussed.    

\subsection{Most likely candidates}
%-------------------------------------------------------------- 
\subsubsection{SC4 53898 and SC9 175336}
These two binaries may serve as the best candidates because of the presence 
of complete eclipses, high brightness
(especially SC9 175336), modest proximity effects allowing for
reasonable determination of the third light contribution to the system and at
last, small residuals. In both cases the primary is slightly cooler and considerably 
larger than the secondary component. As a result there is a quite high
luminosity ratio between components $l_2/l_1 \sim 3$ (see Table~\ref{tab:solA}) which 
may by regarded as the only disadvantage of using these systems.
The expected velocity semiamplitudes should be $K_1 \sim 250$ km/s and $K_2 \sim
190$ km/s for SC4 53898, while $K_1 \sim 220 $ km/s and $K_2 \sim 190$
km/s for SC9 175336. The latter lies near the large H$\,${\footnotesize II}
emission region and OB association -- NGC
371. Both stars were not included by WW1 or WW2 as likely 
distance indicators. 

\subsubsection{SC5 129441 and SC6 67221}
Very high eccentricity and clear proximity effects visible near the periastron 
passage are features of both systems. SC5 129441 consists of two
somewhat evolved B3 stars of equal temperatures. In the second system -- SC6 67221 -- during the 
primary eclipse (phase 1.0) the cooler and probably larger component is eclipsed 
and the strong mutual reflection causes the brightening of the system visible as
a hump around phase 1.0. Both systems may serve as ideal testers of mutual
reflection treatment used in modern light curve synthesis programs. Indeed,
the presence of small residuals visible near phase 1.1 in SC5 129441 and near the primary 
eclipse in SC6 67221 may (Fig.~\ref{fig:fitA}) suggest that
simple reflection used in the analysis (see \S 3.1) might be 
too crude a simplification. Unfortunately the use of the detailed
reflection (which should be more appropriate for these binaries)
together with the eccentric orbits makes WD program practically unsuitable
for analysis due to an enormously large computational time (Wilson 1992). 

Both systems show no substantial third light contribution. 
The expected velocity semiamplitudes should
be $K_1\approx K_2 \approx 140 $ km/s for SC5 129441 and $K_1
\sim 150 $ km/s and $K_2 \sim 140 $ km/s for SC6 67221. 
SC5 129441 was suggested by WW1 to have complete eclipses, but the 
present light curve analysis did not support this thesis. SC6 67221 was not 
included by WW1.  

\subsubsection{SC3 139376}
This binary seems to be the most massive and the most luminous in the sample 
(beside SC9 163575), consisting of two B0 V-IV stars of similar temperature: 
$T \sim 32000$ K. It is the only "classical" well-detached system included in the subset of
likely distance indicators -- the fractional radii of both components are below 0.2.
The residuals of solution are very small and only some larger scatter can be
seen during eclipses due to very fast apsidal motion. 
The rate of the motion is $\dot\omega \approx$ 5\fdg 8/year, so the apsidal period $U$ is
only $\sim 60$ years. Thus the analysis of this system can give parallel 
information about the components and an independent check of stellar
evolution models. The expected velocity semiamplitudes should be
$K_1\sim 190$ km/s and $K_2\sim 180$ km/s. 
  
\subsubsection{SC6 215965}
The brightest binary in the sample ($V=13.94$) with the spectrum B0 V-IV + B1 IV.
Table~\ref{tab:par} suggests that the cooler component is more massive one 
and it is slightly evolved. 
The light curve solution showed that the 
contribution of the third light to this system
might be neglected and photometric parameters such as the temperature ratio 
(surface brightness ratio) could be very accurately determined. 
Also very small residuals suggested that the system had no essential intrinsic 
variability.  This system was yet used by Harries et al.~(2003) for 
a distance determination to SMC giving $m-M=18.83\pm0.15$. The physical parameters
of both components are reasonably close to that presented in Table~\ref{tab:par}. It can
be considered as an independent check of the method given in Section~\ref{physics}
and an additional argument for use this star for an accurate 
individual distance determination.    
WW1 included this binary in their catalog on the basis of having 
complete eclipses but the present analysis (and results of Harries et al.) 
exludes this suggestion.   
%\begin{figure*}
%\begin{minipage}{\linewidth}
%\centering\resizebox{0.8\linewidth}{!}{\includegraphics{smc_uks4.ps}}
%\end{minipage}
%\caption{The position of six eclipsing binaries selected as best distance
%indicators to the Small Magellanic Cloud.}
%\label{fig:cand}
%\end{figure*}

\subsection{SC5 38089 and the candidates suggested recently by Wyithe \& Wilson}
%--------------------------------------------------------------
\subsubsection{SC5 38089}
Harries et al.~(2003) determined absolute dimensions and a distance to this B0V+B0V system: 
$m-M=18.92\pm0.19$. Their light curve solution is essentially the same as the solution reported 
in this paper. However, there is an apparent discrepancy between the observational data on
SC5 38089 and the mass-luminosity relationship shown on Fig~\ref{fig:law}. 
Thus I would not recommended this star for individual distance determination, 
although it is very interested target for spectroscopic investigations in the future.  
Table~\ref{tab:par} gives the estimate of physical parameters assuming that both components 
are normal main sequence stars. The expected $B$ luminosity ratio is very close to unity what 
indeed was reported by Harries et al.

\subsubsection{SC4 103706}
Formally this binary also seems
to have complete eclipses (it was included by WW1 in their subset of 
binaries showing totality) but the very large third light contribution ($l_3
\sim 0.4$) may cause serious problems during modelling of the spectral energy
distribution and thus I decided not to include this star into the
most suitable candidates.  

\subsubsection{SC4 163552}
WW2 included this binary according to the presence of
complete eclipses, but as in the case of SC6 215965,
the light curve analysis contradicted this thesis. 
The third light contribution to this system is sunstantial, at about 26\%.
This binary has small residuals, a very smooth light curve and can
be considered as a "reserve" candidate. The mass ratio and temperature ratio
are very close to unity. The expected velocity amplitude is $K\sim 250$
km/s for both components. 

\subsubsection{SC6 11143}
WW1 classified this system as a well-detached. Indeed, my analysis showed that
the components had the smallest fractional radii of all the
sample (beside SC6 67221) and are normal B2 V stars. However, 
this binary happens to blend with another eclipsing binary (see \S 2.2) 
which contributes about 30\% to the total light of system.
Thus any detailed analysis of SC6 11143 should at the same time take into account 
a careful analysis of the blending binary. Probably this
includes simultaneous solution of multicolour light curves of both binaries 
what is, in principle, possible, but may give ambiguous results. 
 
\subsubsection{SC6 319966}
This binary was included by WW1 according to the presence of
complete eclipses which seems to be incorrect in the view of my analysis. 
This binary had the highest $l_3$ contribution from all the sample --
nearly 50 \% -- which is most probably too large for a detailed modelling of 
spectrophotometry. 
%-------------------------------
\section{Concluding remarks}
%-----------------------------
\subsection{Third light contribution problem}
\label{l3}
The observed difference of the $l_3$ contribution to the total light between
binaries from A and B subsets may be understood as follows.
The components of binaries
from B subset are relatively much closer to each other than those from A
subset and the proximity effects, depending on the gravitational
distortion and overall geometry of the system, are much more distinct. 
Thus proximity effects give us additional constraints to the solution.
When we consider the third light $l_3$ we see
that its influence on the light curve of well-detached systems is
of little importance: equally good solutions can be obtained for 
wide range of $l_3$ values. 

The situation is different for closer systems -- small changes of $l_3$ produce 
noticeably systematic O-C residuals,
especially near quadratures. This is an important reason for preferring such
systems over classical well-detached systems in determination of the
distance using individual eclipsing binaries, especially in the
case of the extragalactic binaries observed usually on a
rich stellar background of a host galaxy. The probability of blending
with another star(s) in such environment is high (crowding effect). 
Also the blend may
be a member of the system itself as at least 30\% of binaries (Batten 1973)
are found to form hierarchical multiple systems of stars. 
%If the blended star (or stars) is a variable we have a somewhat lucky 
%circumstance -- we know about the third light contribution. But if it isn't? 
%What happens if the contribution is of order 5-10 \%?
%Now, the light curve of well-detached
%binary give us no clue about $l_3$. Even if we have other
%possible source of an information e.g. high resolution spectroscopy -- 
%such weak component may escape from an identification (e.g.~by
%line profile changes analysis) and we can
%only put $l_3=0$, which leads us to more or less defective solution.  
Indeed, neglecting of the third
light contribution in the case of HV 2274 and HV 982 may
be one of the prime sources of small differences on the level 1.5$\sigma$ 
between distances derived for three LMC systems as was announced 
by Ribas et al.~(2002).   

\subsection{Fast information about absolute dimensions}
\label{fast}
The idea of deriving the absolute properties of binary stars from knowledge
of their photometric properties alone has long tradition
(e.g.~Gaposchkin 1938, Kopal 1959 - Chapter~7). The methods based on this idea use
somewhat statistical approach: one or both components of a binary conform 
an empirical mass-luminosity relation and the temperature is imposed
by observed spectral type. 
There are two important advantages of
such statistical method 1) it may be used for large samples of binaries and 
2) it gives fast, though approximate physical parameters of binaries. 
Afterward we can choose from the sample most interesting systems and 
investigate them more accurately using spectroscopy, spectrophotometry 
or multicolor photometry. However, past methods neglected the interstellar 
reddening (which influences both the brightness and the colours) and, 
what is more important, the distance to the binary. As a matter of
fact the distance is irrelevant to these methods and thus it can be derived
simply by comparison with the apparent magnitude $m$, even without any radial 
velocity measurements (e.g.~Gaposchkin 1968, Dworak 1974).       
Moreover, they based their reasoning on quite crude spectral type - temperature
calibrations and, of course, as the temperature come in high powers into equations
involved in the method, the obtained parameters were usually very inaccurate.

In present paper this "old-fashioned" method was innovated to
account for the interstellar reddening $E(B-V)$ and recent $B-V$
colour - temperature calibration (Flower 1996). 
It turns out that such approach causes a big improvement in the 
temperature determination and overall physical parameters as was
demonstrated by an accurate "reproducing" of absolute dimensions of three 
LMC eclipsing binaries. However, it was done for a little bit larger
distance to LMC than the distance derived, on average, from these LMC stars. 
A shift of the distance modulus 
$\Delta(m-M)=+0.1$ is quite small and comes probably from some inaccuracies
of Flower's temperature or/and
bolometric corrections calibration for hot stars.

Anyway we should expect
that this kind of method may give fast approximate determination of absolute 
dimensions for many binaries in the Magellanic Clouds, especially as we have large 
archives of binaries discovered by OGLE, MACHO and EROS 
projects at our disposal, and may be used also for eclipsing binaries in other close
galaxies in the near future.    

\section*{Acknowledgments}
%-----------------------------
I am grateful to the anonymous referee for his/her very valuable comments and suggestions. 
I would like to thank Professor Andrzej Udalski for his comments on OGLE
database and Dr T{\~o}nu Viik for reading the manuscript.

\bsp
\label{lastpage}
\end{document}